\documentclass[draft]{agujournal2019}
\usepackage{url} 
\usepackage[inline]{trackchanges} 
\usepackage{soul}
\usepackage{amsmath}
\usepackage{amssymb}
\usepackage{graphicx}
\usepackage{tikz}
\usepackage{subcaption}
\tikzstyle{arrow} = [thick,->,>=stealth]

\draftfalse

\journalname{JGR: Planets}

\begin{document}

\newcommand*\dd{\mathop{}\!\mathrm{d}}
\newcommand{\eqnref}[1]{(\ref{#1})}
\newcommand{\mb}{\mathcal{B}}
\newcommand{\mm}{\mathcal{M}}
\newcommand{\ml}{\mathcal{L}}

\newcommand{\tmm}{\tilde{\mm}}
\newcommand{\mbx}{\mathbf{x}}
\newcommand{\mbs}{\mathbf{s}}
\newcommand{\figref}[1]{Fig. \ref{#1}}
\newcommand{\bs}[1]{\boldsymbol{#1}}
\newcommand{\bx}{\mathbf{x}}
\newcommand{\bn}{\mathbf{n}}
\newcommand{\bss}{\mathbf{s}}
\newcommand{\lmax}{l_{\text{max}}}
\newcommand{\mbf}[1]{\mathbf{#1}}
\newcommand{\hz}{\hat{\zeta}}


\title{Forward and adjoint calculations of gravitational potential in heterogeneous, aspherical planets}

%
%

\authors{Alex D. C. Myhill\affil{1}\thanks{E-mail: adcm2@cam.ac.uk},
Matthew A. Maitra\affil{2},
David Al-Attar\affil{1}}

\affiliation{1}{Department of Earth Sciences, Bullard Laboratories, University of Cambridge}
\affiliation{2}{ETH Zurich}

\correspondingauthor{Alex Myhill}{adcm2@cam.ac.uk}


\begin{keypoints}
\item Computational package for gravitational potential calculation in the referential formulation
\item Forward perturbation theory developed in referential formulation
\item Adjoint theory for gravitational potential sensitivity kernels in the referential formulation
\end{keypoints}

%


\begin{abstract}
We have developed a computational package for the calculation of numerically exact internal and external gravitational potential, its functional derivatives and sensitivity kernels, in an aspherical, heterogeneous planet. We detail our implementation, utilizing a transformation of the Poisson equation into a reference domain, as well as a pseudospectral/spectral element discretisation. The use of the forward solver within the package is demonstrated by calculating the gravitational potential of Phobos with homogeneous and heterogeneous density models. Equations for the first-order perturbation expansion of potential in the referential formulation are found, and the magnitude of the error is quantified based on the exact method. The adjoint Poisson equation is derived, and from it the sensitivity kernels for objective functionals of the potential, which are calculated for Phobos. The expression for perturbations to an objective functional is reduced to a single body and surface integral. Finally, a relation is obtained between the sensitivity kernels which must be satisfied when using a computational domain different to the physical domain.
\end{abstract}

\section*{Plain Language Summary}
We have developed code to find the gravitational potential of a complex body. The code is efficient as a result of formulating the problem in a simpler domain and its usage is demonstrated in calculating the gravitational potential of Phobos\textemdash a moon of Mars. Equations are derived which give the effect of small changes in the planetary shape and the method of solution is demonstrated. Finally, functions of potential are considered and the effect of small changes to the planetary shape on the value of these functions is established.

%
%

\section{Introduction}\label{sec:intro}
\subsection{Preliminaries}
The calculation of Newtonian gravitational potentials in an accurate and efficient manner is important in many areas of planetary, geo- and astrophysics. This calculation is straightforward for bodies which possess a spherical exterior surface, and for which all internal boundaries are concentric spherical shells (termed geometrically spherical herein). The potential in these bodies can be expressed as a sum of spherical harmonics with radially varying coefficients\textemdash yielding ordinary differential equations which can be numerically integrated. Most bodies are not, however, geometrically spherical, and implementing a spherical harmonic expansion is infeasible in the general case, at least at a cursory glance.

There exist a number of ways of solving for the gravitational potential in aspherical bodies. If the deviations from sphericity are small, boundary perturbation theory can be used, see e.g.~\citeA{Dahlen_Tromp_1998}. Typically, this is applied to first order, but it has been developed to higher orders. For instance, \citeA{chambat2005earth} derived expressions for the total perturbation up to second order in topography and lateral density variations. They also concluded that the second order terms were significant enough, within certain geophysical applications, that they should not be disregarded. \citeA{balmino1994gravitational} developed expressions for the exterior gravitational potential of an arbitrarily shaped, homogeneous body, based on spherical harmonics. Aside from the implementation difficulty, this method cannot be used to find the interior potential.~\citeA{barnett1976theoretical} developed an analytical method for arbitrarily shaped homogeneous bodies, using a polyhedral representation of the body shape. Although a powerful technique, the expression for the potential involves singularities in logarithm and arctangent terms, which are difficult to deal with satisfactorily~\cite{barnett1976theoretical}. Furthermore, the surface potential is not calculable. These techniques are all analytical, or semi-analytical, in nature. They also all suffer, in some manner, from difficulties in implementation or generality.

The difficulty in the analytical-type approaches leads to consideration of fully numeric methods. Numerical calculation of gravitational potential would appear straightforward at first\textemdash the Poisson equation is elliptic and generally lends itself well to numerical solution via finite difference or finite element schemes~\cite{knabner2003numerical}. However, the infinite-distance boundary condition poses a significant problem, and techniques to deal with it need to be employed. One option is to employ a very large computational domain, meshing outer space well outside the domain of the planet. This, however, runs into significant computational and accuracy difficulties~\cite{tsynkov1998numerical,gharti2017spectral}. An alternative approach is to employ a Dirichlet-to-Neumann (DtN) map, whereby the solution in the region outside the planet is used to set the boundary conditions on a finite domain, either the boundary of the planet or slightly outside that boundary~\cite{givoli1989finite}. The DtN map, with an exterior spherical harmonic expansion, was combined with a spectral element discretisation of the Earth by~\citeA{chaljub2004spectral}. More recently,~\citeA{van2021modelling} used a similar approach for the perturbed gravitational potential in a seismological setting, defining the Neumann boundary condition at a finite radius exterior to the planet. Alternative approaches to deal with the boundary condition in a spectral element implementation have also been developed with the use of an exterior layer of ``infinite'' elements (see~\citeA{gharti2018spectral,gharti2019spectral,gharti2023spectral}). Fully numeric methods, and spectral element methods specifically, can model arbitrarily complex bodies. However, avoiding convergence issues and calculating to a high spacial precision requires taking a fine discretisation. The correspondingly high number of degrees of freedom results in a very significant computational cost, regardless of the model shape. Furthermore, these techniques generally rely on parallel processing on hundreds of processors, which can be a severe restriction on the workflow. An example of the cost in seismological applications is provided in~\citeA{gharti2023spectral}. This calculation modelled seismic waves accurately to periods of around 20 mHz, corresponding to approximately $\lmax\sim 200$. It took approximately, or slightly over, one CPU-hour to perform a single time step, for which the calculation of the gravitational potential was the most significant cost. It is clear that although fully numeric spectral element methods are very powerful, their computational cost is a significant drawback.

It is unsatisfying, and a significant hindrance, that solving exactly for the gravitational potential has such a disparity in cost between a geometrically spherical planet, and an (even slightly) aspherical one. It was this issue that motivated~\citeA{maitra2019non} (herein MA19) to develop an alternative numerical method which allows one to use a spherical harmonic expansion in geometrically aspherical planets. They introduced a ``reference'' planet which is geometrically spherical, with an associated mapping to the physical planet. The key premise of this method is that simplicity of the Poisson equation can be traded for simplicity of the geometry. One of the main advantages is the scaling of the computational cost. In particular, for spherical planets the cost is similar to that associated with the direct integral techniques. The cost scales, roughly, with the complexity of the planetary geometry, i.e. its deviation from sphericity. 

The discussion of the general Poisson problem thus far has focussed on the forward problem, i.e. calculating the gravitational potential associated with a particular density distribution. Within geophysical or astrophysical problems, potential observations are commonly used to make inferences about the properties (density and shape) of the planet (see e.g.~\citeA{zhdanov2002geophysical}). In this case, being able to solve the general forward problem is insufficient. Within almost all formulations of inverse problems, solving the linearised problem (or calculating functional derivatives of the potential with respect to model parameters) is important~(see e.g. \citeA{backus1967numerical,zhdanov2002geophysical}). Furthermore, many modern approaches to inverse theory rely on the use of sensitivity kernels calculated via the adjoint method~\cite{louis1996approximate,cui2024glad}. The application of these methods within the gravity problem was one of the main motivations for this work. The ultimate goal of the paper is to present a unified and theoretically complete method (and computational package) to address both the forward and inverse Poisson problems within the referential framework.

\subsection{Motivation} \label{sec:motivation}
To complete the introduction we motivate the approach taken within this paper to the forward and inverse Poisson problem. The referential formulation used throughout will be explained in greater detail in Sec.~\ref{sec:referentialformulation}. For now, we content ourselves to describing the problem schematically. The reference problem is fully defined by the reference planet $\tilde{\mathcal{M}}$ with a density $\rho(\mbx)$ and a mapping to the physical planet $\bs{\xi}(\mbx)$, defined throughout $\tilde{\mm}$. Associated with the mapping is the deformation gradient $\mbf{F} = (\bs{\nabla}\bs{\xi})^T$. The gravitational potential that we wish to calculate is $\zeta$, i.e.
\begin{equation}
    \zeta = \hat{\zeta}(\rho,\bs{\xi}, \mbf{F}),
\end{equation}
where $\hat{\zeta}$ is the non-linear operator which maps the parameters $\rho$, $\bs{\xi}$ and $\mbf{F}$ to the potential, defined throughout space. Although $\mbf{F}$ is the gradient of $\bs{\xi}$, it is helpful mathematically, in particular in the inverse problem, to regard it as a separate parameter, which the potential depends upon. To solve the forward problem we need the equation of motion and a method for its numerical discretisation and solution. These are derived and discussed in Sec.~\ref{sec:forward}.

The inverse Poisson problem is the problem of constraining internal structure using gravitational potential data. It is relatively easy to pose, whilst being significantly more complicated to solve. Presuming that we have made measurements of the potential, or some function of the potential, outside the planet, how can we deduce information about the density variation within the planet? We may have, for instance, a model predicated on some assumptions about the interior of the planet, based on other measurements or mineral physics. We then wish to improve our model for the planet based upon the new information, i.e. the measurements of potential. Ultimately, the problem becomes one of finding a model of the planet such that some measure (commonly known as the misfit) of the difference between the calculated and physical observations is minimised. This is a non-linear problem; one method for performing the minimisation is to approach it within a linearised framework. For this, we require the ability to calculate first-order, or linearised, perturbations to the potential, associated with perturbations to model parameters, i.e.
\begin{equation}
    \delta \zeta = D_{\rho} \hz \cdot \delta \rho + D_{\bs{\xi}} \hz \cdot \delta \bs{\xi} + D_{\mbf{F}} \hz \cdot \delta \mbf{F} ,
\end{equation}
where $D_{\rho} \hz$, $D_{\bs{\xi}} \hz$ and $D_{\mbf{F}} \hz$ are the functional derivatives\textemdash operators whose action (represented by the product symbol) maps the perturbation in the parameter of interest to the perturbation in the potential, with the other parameters held constant. Correct to first order, this perturbation will ultimately be given by three separate integrals over the body.

We may also wish to consider scalar functionals of potential, i.e. $Q(\zeta)$. As the potential $\zeta$ depends on the model  parameters $\rho,\bs{\xi}$ and $\mbf{F}$, $Q(\zeta)$ implicitly does too. It is useful to define the reduced objective functional 
\begin{equation}
    \hat{Q}(\rho,\bs{\xi},\mbf{F}) \equiv Q[\hz(\rho,\bs{\xi},\mbf{F})],
\end{equation}
which gives the value of $Q(\zeta)$ which occurs for a potential $\zeta$ corresponding to the parameters $(\rho,\bs{\xi},\mbf{F})$. The perturbation to $Q$ can be written as
\begin{align}
    \delta \hat{Q} &= \left(DQ, D_{\rho}\hz\cdot\delta \rho + D_{\bs{\xi}}\hz\cdot\delta \bs\xi + D_{\mbf{F}}\hz\cdot\delta \mbf{F}\right), \nonumber\\
    & = \left((D_{\rho}\hz)^\dagger DQ(\zeta), \delta \rho\right) + \left((D_{\bs{\xi}}\hz)^\dagger DQ(\zeta), \delta \bs{\xi}\right)+ \left((D_{\mbf{F}}\hz)^\dagger DQ(\zeta), \delta \mbf{F}\right),
\end{align}
where a superscripted dagger indicates the adjoint of the operator. The terms $(D_{\rho}\hz)^\dagger DQ(\zeta)$, $(D_{\bs{\xi}}\hz)^\dagger Q(\zeta)$ and $(D_{\mbf{F}}\hz)^\dagger DQ(\zeta)$ are the sensitivity kernels of $\hat{Q}(\zeta)$ with respect to density, the mapping and the deformation gradient respectively. We have introduced the inner product $(\cdot,\cdot)$, defined as
\begin{equation}
    (\psi,\chi)  = \int_{\mathcal{M}} \left<\psi, \chi\right> \dd V,
\end{equation}
where $\left<\cdot,\cdot\right>$ is the Euclidean scalar product appropriate to the scalar, vector or tensor variables $\psi$ and $\chi$: for scalars $\left<\psi, \chi\right> = \psi \chi$, for vectors $\left<\bs{\psi}, \bs{\chi}\right> = \sum_i \psi_i \chi_i$, whilst for tensors $\left<\bs{\psi}, \bs{\chi}\right> = \psi_{ij} \chi_{ij}$. As $\hat{\zeta}$ is nonlinear (and $Q(\zeta)$ may be too), the sensitivity kernels are not linearly related to those of $\zeta$. Instead, we need to use the constrained Lagrangian technique to derive the so-called adjoint equation, from which the sensitivity kernels can be found.

It is necessary to be able to calculate the functional derivatives of the potential with respect to the three parameters defining the referential model, and sensitivity kernels for scalar functionals of the potential, if one wishes to be able to use the referential formulation within an inverse problem context. More fundamentally, this also ``completes'', in an intellectually satisfying manner, the Poisson problem. In Sec.~\ref{sec:forward} we address the forward problem and discuss in detail the numerical implementation of the solution. In Sec.~\ref{sec:sensitivity} we address the problem of finding the first-order forward derivatives of the potential, whilst in Sec.~\ref{sec:adjoint} we address the problem of finding the sensitivity kernels for an arbitrary functional of potential. In Sec.~\ref{sec:adjoint} we also address the theory of sensitivity kernels within the referential formulation.

\section{Forward modelling} \label{sec:forward}

\subsection{Theory }
We summarise the key details of~MA19, but for a more complete treatment please refer to the original paper. In the interest of simplicity we employ the same notation.

\subsubsection{Poisson's equation and the weak form}
The body, whose gravitational potential we are interested in calculating, occupies $\mathcal{M} \subseteq \mathbb{R}^3$, with a smooth exterior boundary $\partial \mathcal{M}$. The body is allowed to have a finite number of non-interpenetrating layers, with the union of all boundaries and the external boundary given by $\Sigma$. The gravitational potential $\phi$ satisfies the Poisson equation
\begin{equation}
    \nabla^2 \phi = 4 \pi G \varrho, \label{eqn:poissonstrong}
\end{equation}
where $\varrho$ is the density (non-zero only within $\mathcal{M}$), $G$ is the universal gravitational constant and $\nabla^2$ is the Laplacian operator. The boundary and continuity conditions are
\begin{enumerate}
    \item $[\phi]_-^+ = [\left<\hat{\mathbf{n}},\nabla \phi\right>]_-^+ = 0,$ on $\Sigma$,
    \item $\phi \rightarrow 0$ as $||\bx || \rightarrow \infty$,
\end{enumerate}
where $\hat{\bn}$ is the unit normal to the surface $\Sigma$, $[\cdot]_-^+$ is the jump in value across $\Sigma$ in the direction of $\hat{\bn}$, $\left<\cdot,\cdot\right>$ is the Euclidean scalar product and $||\cdot||$ is the associated norm.

As discussed in Sec.~\ref{sec:intro}, one of the main difficulties associated with numerical solutions of the Poisson equation is the infinite distance boundary condition. In the problem we address, however, the density outside of $\mm$ is zero. Consequently, if the potential is known on a spherical surface completely enclosing the body (for which we choose the exterior surface of $\mb = \{\bx \in \mathbb{R}^3| \phantom{=}||\bx|| \leq b\}$, such that $\mm$ is completely contained within $\mb$), then it is trivial to compute the potential outside the surface in terms of spherical harmonics (see e.g.~\citeA{chaljub2004spectral}, or Appendix B of~\citeA{Dahlen_Tromp_1998}). The problem now becomes one of solving~\eqref{eqn:poissonstrong} within $\mb$. Boundary conditions associated with the exterior solution are defined on $\partial\mb$. MA19 use the weak form to define these boundary conditions, introducing the test function $\psi$, defined on $\mb$, multiplying~\eqref{eqn:poissonstrong} by its conjugate, and integrating over the domain, obtaining
\begin{equation}
    \int_{\mb}( \nabla^2 \phi) \overline{\psi} \dd^3 x = 4\pi G \int_{\mm} \varrho \overline{\psi} \dd^3 x.
\end{equation}
Making use of Gauss' theorem (see e.g. Appendix A of~\citeA{Dahlen_Tromp_1998}) 
\begin{equation}
    \int_{\mb} \left<\nabla \phi, \nabla\bar{\psi} \right> \dd^3 \bx - \int_{\partial \mb} \left< \hat{\mathbf{n}},\nabla \phi\right> \dd S
    = -4 \pi G \int_{\mm} \varrho \bar{\psi} \dd^3 \mathbf{x},
\end{equation}
where $\left< \hat{\mathbf{n}},\nabla \phi\right>$ is the normal derivative of the interior solution (i.e. the solution $\phi$ for $r < b$) of potential. All terms involving surface integrals on internal boundaries cancel due to the continuity of the normal derivative. Continuity across the exterior boundary $\partial\mb$ is enforced by expressing the exterior solution in terms of spherical harmonics as
\begin{equation}
    \phi(r,\theta,\varphi) = \sum_{lm} \left(\frac{b}{r}\right)^{l + 1} \phi_{lm}(b) Y_{lm}^0(\theta,\varphi),
\end{equation}
where $Y_{lm}^N$ is the generalised spherical harmonic of degree $l$, order $m$ and upper index $N$ (e.g. Appendix C of~\citeA{Dahlen_Tromp_1998}). The coefficients $\phi_{lm}$ are the (generalised) spherical harmonic coefficients at degree $l$ and order $m$, defined as
\begin{equation}
    \phi_{lm}(b) = \int_{\mathbb{S}^2} \phi(b,\theta,\varphi) \overline{Y_{lm}^0(\theta,\varphi)} \dd S,
\end{equation}
where $\mathbb{S}^2$ is the unit 2-sphere and an overline denotes complex conjugation. The DtN map is now imposed\textemdash enforcing continuity of the normal derivative across $\partial\mb$\textemdash by finding the derivative of the exterior potential on the boundary and substituting it into the surface integral, yielding the weak form
\begin{equation}
    \int_{\mb} \left<\nabla \phi, \nabla\bar{\psi} \right> \dd^3 \bx + \sum_{lm} (l + 1) b \phi_{lm}(b) \bar{\psi}_{lm}(b) 
    = -4 \pi G \int_{\mm} \varrho \bar{\psi} \dd^3 \mathbf{x}, \label{eqn:weaknotransformpot}
\end{equation}
where the spherical harmonic components of $\psi$ are defined similarly to those of $\phi$. This equation must hold for all test functions $\psi$. 

\subsubsection{The referential transformation} \label{sec:referentialformulation}
\begin{figure}
    \centering
    \begin{tikzpicture}
        \centering
        \def\R{2.0};
        \def\RI{0.4 * \R};
        \def\RO{0.55 * \R};
        \def\RB{1.3*\R};
        \def\SP{1.5};
        \def\SD{0.2};
        \draw [black, ultra thick, dashed, fill =blue!50!white, domain=0:360, samples=60]
        plot ({\RB* cos(\x)}, { \RB * sin(\x)} );
    
        \draw [black, ultra thick,  fill = orange!50!white, domain=0:360, samples=60]
        plot ({\R* cos(\x)}, {\R*(sin(\x) - 0.1 * cos(\x) + 0.05 * cos(6 * \x))} );
    
        \draw [black,   ultra thick, fill=yellow!50!white,  domain=0:360, samples=60]
        plot ({\RI * cos(\x)}, {\RI * (sin(\x) + 0.5 * cos(\x))} );

        \draw[black, ultra thick, dashed, fill =blue!50!white] (-2 * \RB - 2*\SP,0) circle(\RB);
        \draw[black, ultra thick,  fill = orange!50!white] (-2 * \RB - 2*\SP,0) circle(\R);
        \draw[black, ultra thick,  fill = yellow!50!white] (-2 * \RB - 2*\SP,0) circle(\RI);

        \draw[->,double, ultra thick, line width=2.2pt] (-\RB-2*\SP+0.5*\SD *\SP,0) -- ++ ( 2.0*\SP -   \SD * \SP,0) node[black, midway, above=0.5cm, align=center]{\large Mapping $\bs{\xi}(\cdot)$};
    
        \node[black, ultra thick] at (-2*\RB - 2 * \SP, \RB + 0.5) {\large Reference planet};
        \node[black, ultra thick] at (0, \RB + 0.5) {\large Physical planet};

        \def\XC{-2 * \RB - 2*\SP};
        \def\OX{1.5*\RI};
        \draw[->,ultra thick] (\XC + \OX,\OX) to[out=25,in=155] (\OX,\OX);
        \filldraw (\XC + \OX,\OX) circle(2pt) node[align=right, below] {\large $\bx$};
        \filldraw (\OX,\OX) circle(2pt) node[align=right, below] {\large $\bs{\xi}(\bx)$};

    \end{tikzpicture}
    \caption{A schematic depicting the mapping from a geometrically spherical reference planet to a physical planet, each comprised of two layers. The dashed line represents the surface of $\mb$, whilst the outer and inner solid lines represent the exterior of the planet and the boundary between the two layers (shown in different colours) respectively. The reference planet has geometrically spherical boundaries whilst the physical planet is arbitrary.} 
    \label{fig:mapping}
\end{figure}
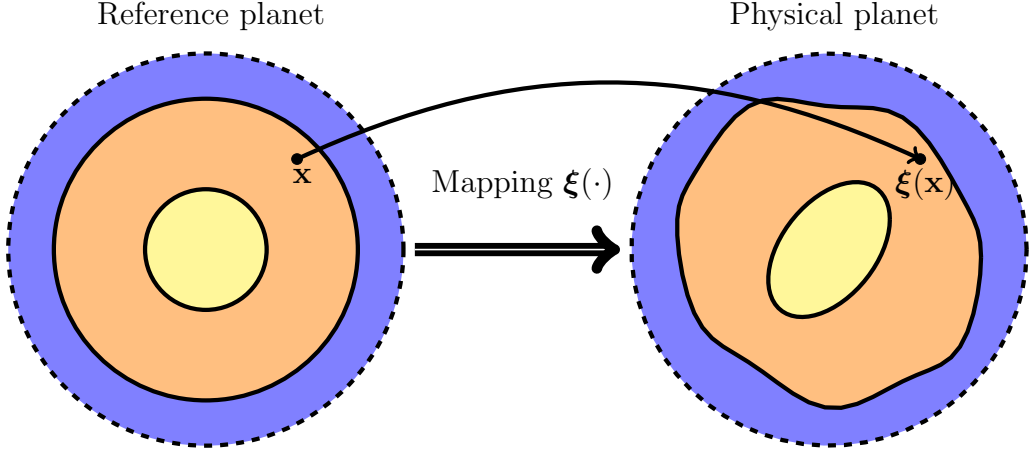

The weak form given by~\eqref{eqn:weaknotransformpot} is suitable for numerical discretisation. However, in this form, it is difficult to use spherical harmonic expansions, due to continuity conditions on internal aspherical boundaries~(MA19). MA19 consider a diffeomorphism (an invertible function mapping one domain to another, such that the function and its inverse are continuous) $\bs{\xi}: \mb \rightarrow \mb$ (see~\figref{fig:mapping}), with the properties:
\begin{enumerate}
    \item On the boundary $\partial \mb$ it is the identity mapping;
    \item The inverse image $\tilde{\mm} = \bs{\xi}^{-1}(\mm)$ is a ball with centre coincident with the centre of $\mb$;
    \item The inverse image of $\tilde{\Sigma} = \bs{\xi}^{-1}(\Sigma)$ is a set of concentric spherical shells in $\tilde{\mm}$. 
\end{enumerate}
The inverse image of $\mm$ under the mapping, $\tilde{\mm}$, is known as the referential planet. The deformation gradient $\mathbf{F}$, is defined as 
\begin{equation}
    F_{ij} = \frac{\partial \xi_i}{\partial x_j}.
\end{equation}
Its Jacobian, $J = \det \mbf{F}$, represents the volume mapping induced by $\bs{\xi}$. The Cauchy-Green deformation tensor, which may be recognisable to readers familiar with elasticity, is given as $\mbf{C} = \mbf{F}^T \mbf{F}$. We also define the tensor $\mbf{a} = J \mbf{C}^{-1}$. 

To transform the equation into one defined on the reference planet, we define the referential potential field as
\begin{equation}
    \zeta(\mathbf{x}) = (\phi \circ \bs{\xi})(\bx) := \phi[\bs{\xi}(\bx)].
\end{equation}
Using $\bs{\xi}$ to transform~\eqnref{eqn:weaknotransformpot} then yields
\begin{equation}
    \int_{\mb} \left<\mbf{a} \nabla \zeta, \nabla\bar{\chi} \right> \dd^3 \mathbf{x} + \sum_{lm} (l + 1) b \zeta_{lm}(b) \bar{\chi}_{lm}(b) = -4 \pi G \int_{\tilde{\mm}} \rho \bar{\chi} \dd^3 \mathbf{x}, \label{eqn:weakpot0}
\end{equation}
where $\chi(\bx) = (\psi \circ \boldsymbol{\xi})(\bx)$ is the transformed test function. The referential density is defined as $\rho(\bx) = J(\bx) (\varrho \circ \boldsymbol{\xi})(\bx)$. As $\psi$ was an arbitrary test function in $\mathcal{M}$, so is $\chi$ in $\tilde{\mm}$. Thus,~\eqref{eqn:weakpot0} must hold for all test functions $\chi$. This form of Poisson's equation is more complicated than the earlier weak form. However, the geometry is simpler, and it is now possible to employ spherical harmonic expansion methods. It should be emphasised that the only requirement on $\mb$ is that it totally encloses the planet within it (and is spherical). It is, however, purely computational in nature and does not have any physical significance.

\subsection{Numerical implementation}
\subsubsection{Generalised spherical harmonics}
Although some readers may be familiar with generalised spherical harmonics (GSHs), we feel it is worth briefly outlining some of their properties and the manner in which we use them\textemdash further information can be found in~\citeA{phinney1973representation} and~\citeA{Dahlen_Tromp_1998}. GSHs are a convenient, and indeed ``natural'', way of representing tensors possessing spherical symmetry~\cite{burridge1969spherically}. The GSH expansion for a tensor of rank $n$ is 
\begin{equation}
    \mbf{T}= \sum_{l = 0}^\infty \sum_{m= -l}^l \sum_{\{\alpha\}}T_{lm}^{\alpha_1 ... \alpha_n} (r) Y_{lm}^N(\theta,\varphi) \hat{\mbf{e}}_{\alpha_1} \otimes \hat{\mbf{e}}_{\alpha_2} \otimes .. \otimes \hat{\mbf{e}}_{\alpha_n},
\end{equation}
where we write $\otimes$ for tensor products and define $Y_{lm}^N(\theta,\varphi)$ as the GSH of degree $l$, order $m$ and upper index $N = \sum_{i} \alpha_i$. The canonical basis vectors $\hat{\mbf{e}}_{\alpha}$ are defined as the eigenvectors of the infinitesimal rotation matrix corresponding to rotations about the local $\hat{\mbf{r}}$-axis~\cite{burridge1969spherically,phinney1973representation}, given by
\begin{align}
    \hat{\mbf{e}}_{-} & = \frac{1}{\sqrt{2}}(\hat{\bs{\theta}} - i\hat{\bs{\varphi}}), & \hat{\mbf{e}}_0 & = \hat{\mbf{r}}, & \hat{\mbf{e}}_{+} & = -\frac{1}{\sqrt{2}}(\hat{\bs{\theta}} + i\hat{\bs{\varphi}}).
\end{align}
The canonical basis vectors satisfy the orthogonality property $\hat{\mbf{e}}_{\alpha} \cdot \overline{\hat{\mbf{e}}_{\beta}} = \delta_{\alpha, \beta}$. $T_{lm}^{\alpha_1 ... \alpha_n} (r) $ is the expansion coefficient for the basis tensor $Y_{lm}^{N}\hat{\mbf{e}}_{\alpha_1} \otimes ... \otimes \hat{\mbf{e}}_{\alpha_n}$. The GSHs themselves take the form $Y_{lm}^N(\theta,\varphi) = X_{lm}^N(\theta)e^{im\varphi}$, where $X_{lm}^N(\theta)$ is a generalised Legendre function (see~\citeA{Dahlen_Tromp_1998} Appendix C.4). It should be noted that the generalised Legendre functions are, up to normalisation, equivalent to values of the Wigner d-matrix~\cite{phinney1973representation}, and they can be calculated using stable recursive schemes. Finally, the GSH orthogonality relation is given by
\begin{equation}
    \int_{\mathbb{S}^2}  \overline{Y_{lm}^N} Y_{l'm'}^{N} \dd S = \delta_{ll'}\delta_{mm'}.
\end{equation}
The orthogonality relation ensures that three-dimensional integrals can be reduced to a set of radial integrals of GSH coefficients. The representation of gradients of tensor fields is straightforward using GSH expansions. In the case of a scalar field given by $f = \sum_{lm} f_{lm} Y_{lm}^0$ the gradient can be written
\begin{equation}
    \nabla f = \sum_{lm} \left\{\Omega_{l}^{0} r^{-1} f_{lm} [Y_{lm}^{-}\hat{\mbf{e}}_{-}+Y_{lm}^{+}\hat{\mbf{e}}_{+}] + \partial_r f_{lm} Y_{lm}^0 \hat{\mbf{e}}_0\right\}, \label{eqn:gradgsh}
\end{equation}
where $\Omega_{l}^{N} = \sqrt{(l + N)(l - N+1)/2}$.
This has important implications numerically. In particular, if one has a spherical harmonic expansion for a scalar field (for instance the potential), then the gradient of that potential can be expressed exactly using a GSH expansion. As the values of the GSH can be pre-calculated this leads to significant efficiency gains. 

To use GSHs numerically, we need to consider how they will be employed. We wish to be able to transform between fields defined on a spatial grid and fields defined in terms of GSH expansion coefficients. The transformation from physical space to GSHs is
\begin{equation}
    f_{lm}^N = \int_{\mathbb{S}^2} X_{lm}^N(\theta) e^{-im\varphi} f(\theta,\varphi)\dd S = \int_{0}^{\pi}  \sin \theta X_{lm}^N(\theta) \dd \theta \int_{0}^{2\pi} e^{-im\varphi} f(\theta,\varphi) \dd \varphi.
\end{equation}
The integral over $\varphi$ can be rapidly evaluated using a Fast Fourier Transform (FFT), if the discretisation in $\varphi$ is chosen to be evenly spaced. The discretisation in $\theta$ is chosen to be the Gauss-Legendre nodes corresponding to a domain of $[0,\pi]$\textemdash integrals over $\theta$ are evaluated using Gauss-Legendre quadrature. \citeA{driscoll1994computing} developed a method which has superior theoretical scaling when integrating Legendre functions over $\theta$. However, it has not been implemented for GSHs, the cost of the inverse transformation is the same, and it shows no improvement in the cost of the forward transformation for small $l$~\cite{lesur1999evaluation}. 

The choice of storage order for fields defined in GSH coefficients is important for computational speed. The FFT for the longitudinal integral is fastest if the components are stored in strictly increasing $m$ for the same $l$. This order ensures that the coefficients are accessed by the FFT continuously, maximising cache usage. The storage order is therefore chosen to be
\begin{equation}
    \bx = \begin{bmatrix}
        f_{00}^N &
        f_{1-1}^N &
        f_{10}^N &
        f_{11}^N &
        f_{2-2}^N &
        \hdots &
        f_{\lmax \lmax}^N
    \end{bmatrix}^T.
\end{equation}

\subsubsection{Radial discretisation\textendash spectral elements}
The spherical harmonic expansion of the potential is given by
\begin{equation}
    \zeta(r,\theta,\varphi) = \sum_{lm}\zeta_{lm}(r)Y_{lm}^0(\theta,\varphi).
\end{equation}
MA19 chose to use a spectral element basis for the radially varying coefficients $\zeta_{lm}$. This method has found popularity in the seismology community (e.g.~\citeA{komatitsch1999introduction}) as well as in computational fluid dynamics (e.g.~\citeA{patera1984spectral}). However, it may not be as well known in other disciplines, and a slight elaboration on its theoretical basis and implementation is worthwhile.

Spectral element methods are a subset of finite element methods~\cite{igel2017computational}. A finite element method meshes the domain of the problem into a set of subdomains, known as elements. A set of basis functions, whose support is confined to individual elements, is then defined, based on the mesh~\cite{li2017numerical}. For a problem in the weak formulation, a matrix equation can then be obtained using the Galerkin principle, by requiring that the basis for the expansion of the function is the same as the basis for the test functions~\cite{igel2017computational}. A spectral element method is a finite element method using a set of orthogonal polynomials as the basis within each element~\cite{patera1984spectral}\textemdash combining the computational flexibility and parallelisability of a finite element method with the convergence characteristics of a spectral method~\cite{igel2017computational}. 

We restrict our discussion henceforth to the comparatively simple case of a one-dimensional problem (as we are only interested in the radial expansion). Experience has shown that Lagrange polynomials, defined on a set of nodes, provide a good basis set. The nodes used are the Gauss-Lobatto-Legendre collocation points (appropriately mapped from $[-1,1]$ to the element's boundaries). Numerical evaluation of the integrals within the weak form becomes trivial, as one can employ Gauss-Lobatto quadrature. Although not exact, this is done in almost all practical implementations~\cite{li2017numerical}

To formalise this discussion, we label the p-th Lagrange polynomial in the n-th element $h_{np}$. The radii defining the boundaries between elements are $r_{n}$, i.e. the upper radius of the n-th element is $r_n$. The radial expansion of the $lm$ spherical harmonic component is given by
\begin{equation}
    \zeta_{lm}(r) = \sum_{n}\sum_{p}\zeta_{lmnp}h_{np}(r).
\end{equation}
The support of each basis function is the element, i.e. $h_{np}(r) = 0$ if $r\notin [r_{n-1},r_n]$. The above expression for the radial potential (or its derivative) is, therefore, evaluable with only one sum over $p$. The test functions are the basis functions used, i.e. $\chi_{lmnp} = h_{np}(r) Y_{lm}(\theta,\varphi)$. 

Continuity of the potential needs to be enforced at boundaries between elements. This is achieved by specifying that the coefficients of the Lagrange polynomials on the edge of neighbouring elements is the same. The contributions to the matrices from the two different elements, for that coefficient, are summed together~\cite{igel2017computational}.

\subsubsection{The formal matrix-vector problem and the matrix-free solution}
The weak form is most usefully represented with an operator notation. In particular the equation becomes
\begin{equation}
    \mathcal{A}(\chi,\zeta) = f(\chi), \label{eqn:opform}
\end{equation}
where the sesquilinear form $\mathcal{A}(\cdot,\cdot)$ is defined for arbitrary functions $\chi$ and $\zeta$
\begin{equation}
    \mathcal{A}(\chi,\zeta) = \int_{\mb} \left<\mbf{a} \nabla \zeta, \nabla\bar{\chi} \right> \dd^3 \mathbf{x} + \sum_{lm} (l + 1) b \zeta_{lm}(b) \bar{\chi}_{lm}(b), 
\end{equation}
and the functional $f(\cdot)$ is defined for arbitrary function $\chi$ as
\begin{equation}
    f(\chi) = -4 \pi G \int_{\tilde{\mm}} \rho \bar{\chi} \dd^3 \mathbf{x}.
\end{equation}
The Galerkin principle stipulates that~\eqref{eqn:opform} must hold for all test functions $\chi_{i}$, where $\chi_i$ is the $i$-th basis function and $i\equiv \{lmnp\}$. Substituting into the weak form $\chi = \chi_{i}$, expanding the potential as $\zeta = \sum_{i'}\zeta_{i'} \chi_{i'}$ and using the sesquilinearity of $\mathcal{A}$ we obtain
\begin{equation}
    \mathcal{A}\left(\chi_{i},\zeta\right) = \sum_{i'} \mathcal{A}(\chi_{i}, \chi_{i'}) \zeta_{i'} = f(\chi_i).
\end{equation}
We define the matrix $\mbf{A}$ as $A_{i; i'} = \mathcal{A}(\chi_{i}, \chi_{i'})$, the $i$-th component of the vectors $\bs{\zeta}$ and $\mbf{f}$ as $\zeta_i$ and $f_i = f(\chi_i)$ respectively. We must now solve the matrix-vector problem $\mbf{A}\bs{\zeta} = \mbf{f}$, where the matrix $\mbf{A}$ is Hermitian, which follows from the Hermiticity of the weak form. The maximum degree spherical harmonic at which the expansion is truncated is chosen pragmatically, based on the maximum structural degree in the problem.

The matrix $\mbf{A}$ comprises large, dense sub-blocks and it is not uncommon for it to exceed $10^5 \times 10^5$ in size. Furthermore, evaluation of $\mathcal{A}(\cdot,\cdot)$ is not trivial and explicit calculation of $\mbf{A}$ would be very expensive. Consequently, it is computationally prudent to use iterative solvers\textemdash where an approximate solution is improved successively. We confine our discussion to Krylov subspace methods, which are a type of projection method~\cite{saad2003iterative}. In general, to solve $\mbf{A} \bs{\zeta} = \mbf{f}$, they require an initial guess $\bs{\zeta}_0$ and the ability to pre-multiply by $\mbf{A}$, i.e. to evaluate $\mbf{b} = \mbf{A} \bs{\zeta}$. To solve $\mbf{A} \bs{\zeta} = \mbf{f}$ iteratively, we do not need to assemble $\mbf{A}$. 

There are two questions that naturally arise:
\begin{enumerate}
    \item Is this approach necessarily faster?
    \item How well controlled are the errors?
\end{enumerate}
The key to getting a satisfactory answer to both of these questions lies in pre-conditioning. Pre-conditioning refers to finding a matrix $\mbf{M} \approx \mbf{A}^{-1}$~\cite{saad2003iterative}. The matrix $\mbf{A}_0$ associated with the identity mapping (i.e. $\bs{\xi} = \mbf{x}$), is sparse, has well-known structure and a cheap Cholesky decomposition\textemdash its inverse has proven to be a good pre-conditioner. This allows us to address the two questions raised above:
\begin{enumerate}
    \item Employing $\mbf{A}_0^{-1}$ as a pre-conditioner, the iterative scheme converges within a few tens of iterations at most. Consequently, the number of evaluations of $\mathcal{A}$ is significantly lower than would be required to assemble $\mbf{A}$. The computational savings associated with not solving the large dense system are even more significant.
    \item The approximation error (as measured by the residual) is controlled by the user. Increasing the number of iterations will reduce the error. Practically, the scheme converges to close to machine precision. This is most likely comparable to the error that would be incurred due to round-off if $\mbf{A}$ was assembled and decomposed.
\end{enumerate}

\subsubsection{The evaluation of the sesquilinear form}
Evaluating the sesquilinear form $\mathcal{A}(\cdot,\cdot)$ accurately and efficiently is the remaining critical step. The pseudospectral approach of MA19 is used, i.e. operations are performed in the space within which they are easiest (and most accurate). Multiplications (such as $\mbf{a} \bs{\nabla} \zeta$) are performed in physical space, whilst derivatives are evaluated using generalised spherical harmonic transforms and the spectral element basis. The procedure for evaluating $\mbf{A}\bs{\zeta}$ is as follows
\begin{enumerate}
    \item Find the GSH components of $\mbf{w} = \bs{\nabla} \zeta$ using~\eqref{eqn:gradgsh};
    \item Perform an inverse GSHT of $\mbf{w}$, obtaining its values on the spatial grid;
    \item Pre-multiply by $\mbf{a}$ to form the vector $\mbf{q} = \mbf{a} \mbf{w}$ at each spatial point;
    \item Perform a GSHT of $\mbf{q}$ to obtain its GSH expansion coefficients;
    \item Obtain the final result by performing an integral against the test functions $\chi_{lmnp}=h_{np} Y_{lm}$. To evaluate this we have
    \begin{align}
        \mathcal{A}(\chi_{lmnp},\zeta) & = \int_{\mb} \left<\mbf{a} \nabla \zeta, \nabla\bar{\chi} \right> \dd^3 \mathbf{x} + \sum_{l'm'} (l' + 1) b \zeta_{l'm'}(b) \bar{\chi}_{l'm'}(b), \nonumber \\
        & = \int_{\mb} \left<\mbf{q}, \overline{\nabla(h_{np} Y_{lm})} \right> \dd^3 \mathbf{x} +  (l + 1) b \zeta_{lm}(b) h_{np}(b).
            \end{align}
    This is simplified using the canonical basis orthogonality $\hat{\mbf{e}}_{\alpha} \cdot \overline{\hat{\mbf{e}}_{\beta}} = \delta_{\alpha, \beta}$ and the GSH orthogonality condition, ie 
    \begin{align}
        \mathcal{A}(\chi_{lmnp},\zeta) & = \int_{0}^{b}  r^2 \dd r \int_{\mathbb{S}^2} \left[\Omega_{l}^{0} r^{-1} h_{np} (q^{-}  \overline{Y_{lm}^{-}}+ q^{+} \overline{Y_{lm}^{+}}) + q^0 \partial_r h_{np} \overline{Y_{lm}^0}\right] \dd S , \nonumber\\
        & \phantom{=} +  (l + 1) b \zeta_{lm}(b) h_{np}(b), \nonumber \\
        & = \int_{0}^{b}  \left[r^2 \dot{h}_{np} q_{lm}^0 + \Omega_l^0 r h_{np}(q_{lm}^+ + q_{lm}^-)\right]\dd r +  (l + 1) b \zeta_{lm}(b) h_{np}(b),
    \end{align}
    where $q_{lm}^{\alpha}$ is the $\alpha$-th GSH coefficient for $\mbf{q}$ and $\dot{h}_{np} = \partial_r h_{np}$. Using Gauss-Lobatto quadrature and the fact that $h_{np}(r)$ has support only in the $n$-th element, the integral simplifies to
    \begin{align}
        \mathcal{A}(\chi_{lmnp},\zeta) & =  J_n\Omega_{l}^0 r_{np} w_{p} [q_{lm}^+(r_{np}) + q_{lm}^-(r_{np})] \nonumber\\
        & \phantom{=} + J_n \sum_{p'} w_{p'} r_{np'}^2 q_{lm}^0(r_{np'})\dot{h}_{np}(r_{np'})  \nonumber\\ 
        & \phantom{=} + (l + 1) b \zeta_{lm}(b) \delta_{nN}\delta_{pP},
    \end{align}
    where $w_{p}$ is the $p$-th weight in the Gauss-Lobatto quadrature scheme; $r_{np}$ is the radius of the $p$-th node in the $n$-th element; $J_n$ is the Jacobian of the transformation between the interval $[-1,1]$ and $[r_{n-1},r_n]$, given by $J_n = (r_n - r_{n-1})/2$; $N$ is the outermost element and $P$ is the Lagrange polynomial corresponding to the upper node. The derivatives of the Lagrange polynomials defined on the interval $[-1,1]$ are stored and the radial derivatives $\dot{h}_{np}(r_{np'})$, in the interval $[r_{n-1},r_n]$, are calculated from these by dividing by the Jacobian $J_n$. The integral is straightforwardly, and rapidly, evaluated.
\end{enumerate}

To store the coefficients $\zeta_{lmnp}$, we have chosen to order firstly via element and node and then via spherical harmonic coefficient, i.e. all coefficients at the same radius are stored contiguously, $[x_{np}]_{lm} = \zeta_{lmnp}$. The order of the spherical harmonic coefficients was chosen to be from $0$ to $l_{max}$, and from $m = -l\rightarrow l$. These changes were made to align with the optimal storage of components for the GSH transforms, as discussed above. 

\subsubsection{The mapping between the referential and physical planets}
Thus far we have not considered the form, or properties, of the mapping $\bs{\xi}(\mbf{x})$. MA19 discusses only radial mappings, i.e. mappings for which $\bs{\xi}(\mbf{x}) = [x + h(\mbf{x})] \hat{\mbf{r}}$. Pragmatically, there appears little benefit, for most applications, in considering more complicated mappings\textemdash both computational cost and accuracy are reasonable using only a radial mapping. 

The choice of radial mapping (and referential planet) can vary between different planetary geometries. The only requirement on the referential planet is that it is geometrically spherical. Nonetheless, it is sensible to make the referential planet ``close'' to the physical planet in some sense\textemdash the pre-conditioning assumes an identity mapping, and should be more efficient the closer to identity the actual mapping is. One method for achieving this is to choose the boundary radii within the referential planet to be the average radii of the boundaries on the physical planet and to have a linear interpolation at each set of angles $\{\theta,\varphi\}$. If the lower and upper radii of a particular layer are $r_{n-1}$ and $r_n$ in the referential planet and $p_{n-1}(\theta,\varphi)$ and $p_n(\theta,\varphi)$ in the physical planet then the mapping is
\begin{equation}
    \bs{\xi}(\mbf{x})  = \left[p_{n-1}(\theta,\varphi) + \frac{x - r_{n-1}}{r_n - r_{n-1}} (p_{n}(\theta,\varphi) - p_{n-1}(\theta,\varphi))\right] \hat{\mbf{r}} , 
\end{equation}
which implies that 
\begin{equation}
     h(\mbf{x}) = \left\{x + \Delta_n + \frac{\Delta_{n+1} - \Delta_n}{r_n - r_{n-1}}(x - r_{n-1})\right\}\hat{\mbf{r}} , 
\end{equation}
where we have written $\Delta_i = p_i - r_i$, dropped $(\theta,\varphi)$ for convenience and term $h(\mbf{x})$ the radial displacement. Having obtained $h(\theta,\varphi)$ the calculation of $\mbf{a}$ is straightforward using the results of Appendix B of~MA19.

There are numerical issues associated with the evaluation of $\mbf{a}$ near the origin. Using the above linear interpolation scheme, the radial displacement in the innermost layer will be given by 
\begin{equation}
    h(\mbx) = \left[x + \frac{\Delta_1}{r_1}\right],
\end{equation}
where $\Delta_1$ and $r_1$ refer to the topography and average radius of the first boundary. However, the presence of a non-zero radial displacement close to the origin could result in numerical instability due to division by factors close to zero. It is preferable to take a radius beneath which the displacement is chosen to be zero. We have not, however, found issues with using a linear interpolation throughout the innermost layer.

It should be noted that radial mappings are not sufficiently general to calculate the potential in more complicated geometries that are still diffeomorphic to a ball, for example the synestias of~\citeA{lock2017structure}. However, this extension is not the focus of this paper and finding a more general method for obtaining the mapping is left to future work.

\subsection{Phobos}

\subsubsection{Forward calculation}
\begin{figure}
    \includegraphics[width=\textwidth,trim={4cm 1cm 7cm 2cm},clip]{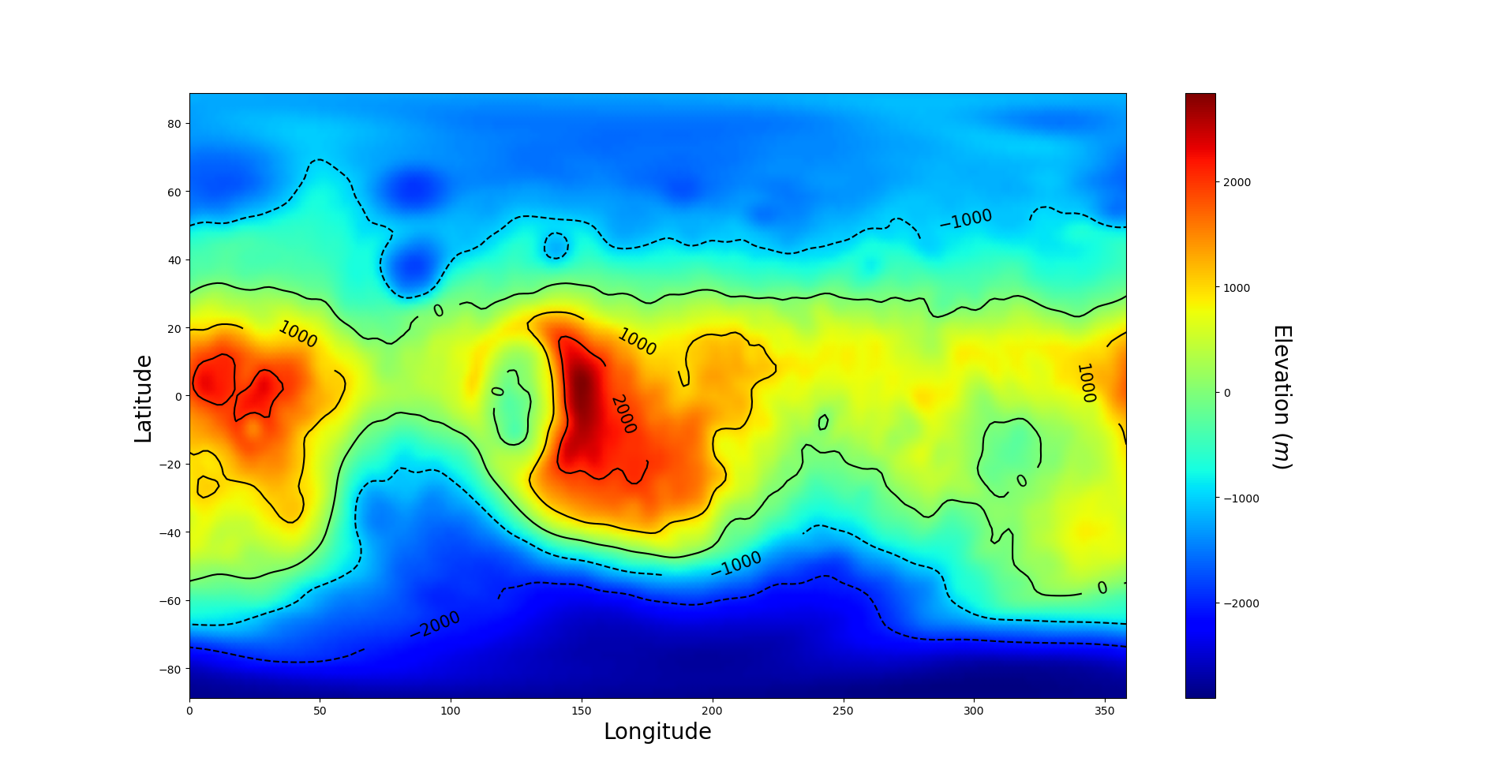}
    \caption{Surface topography of Phobos. Contours are given every 1000 m and the topography is plotted with respect to a sphere of radius $11.1$ km.}
    \label{fig:phobostopography}
\end{figure}

Phobos is the larger of Mars' two moons. It is a small moon that has a peculiar shape, being approximately a triaxial ellipsoid, with radii of 13.0 km, 11.4 km and 9.1 km and a bulk density of 1.86 gcm${}^{-3}$, likely indicating significant porosity~\cite{willner2010phobos,willner2014phobos}. The presence of several large impact craters further complicates calculation of its gravitational potential. The best estimate of its shape is from~\citeA{willner2014phobos}, who published a model of its surface topography expanded in spherical harmonics up to degree 45. A map of its surface elevation with respect to a sphere of radius $11.1$ km is shown in~\figref{fig:phobostopography}. The internal structure of Phobos is not well known, and it is likely that it is a rubble pile~\cite{willner2014phobos,dmitrovskii2022constraints}. Gravitational potential modelling of Phobos as a continuous body is nonetheless useful when investigating its internal structure~\cite{shi2012working,dmitrovskii2022constraints}. 

The referential model is chosen to be a geometric sphere with the average radius of Phobos. To obtain a radial mapping we choose the topography at the surface and linearly interpolate through all except the innermost layer. The radial mapping is then given by 
\begin{equation}
    h(r,\theta,\varphi) = \frac{r}{r_{\text{av}}}\sum_{l=1}^{l_{\text{max}}} \sum_{m = -l}^{l} h_{lm} Y_{lm}, 
\end{equation}
where $h_{lm}$ is the coefficient of the degree $l$, order $m$ spherical harmonic in the expansion of Phobos' surface topography, $r_{\text{av}}$ is its average radius and $l_{\text{max}}$ is the maximum degree of the calculation (with $h_{lm} = 0$ if the degree is greater than the maximum degree of the topography). The maximum degree was chosen for the calculation to be $100$, with $10$ internal elements and the order of the spectral element method was chosen to be $5$. This corresponds to node spacings of approximately $0.01 r_{\text{av}}$. The radial mapping is then evaluated at each point on the grid. The Jacobian, deformation gradient and $\mbf{a}$ tensor are obtained using GSHTs. Finally, the referential density is specified at each point by dividing the physical density by the Jacobian. The calculated volume of Phobos using the referential model defined above is 5741.48 km${}^3$, which is very similar to the $5741.5\pm35$ km${}^3$ calculated by~\citeA{willner2014phobos} using discrete volume elements.

\begin{figure}
    \includegraphics[width=\columnwidth,trim={4cm 1cm 7cm 2cm},clip]{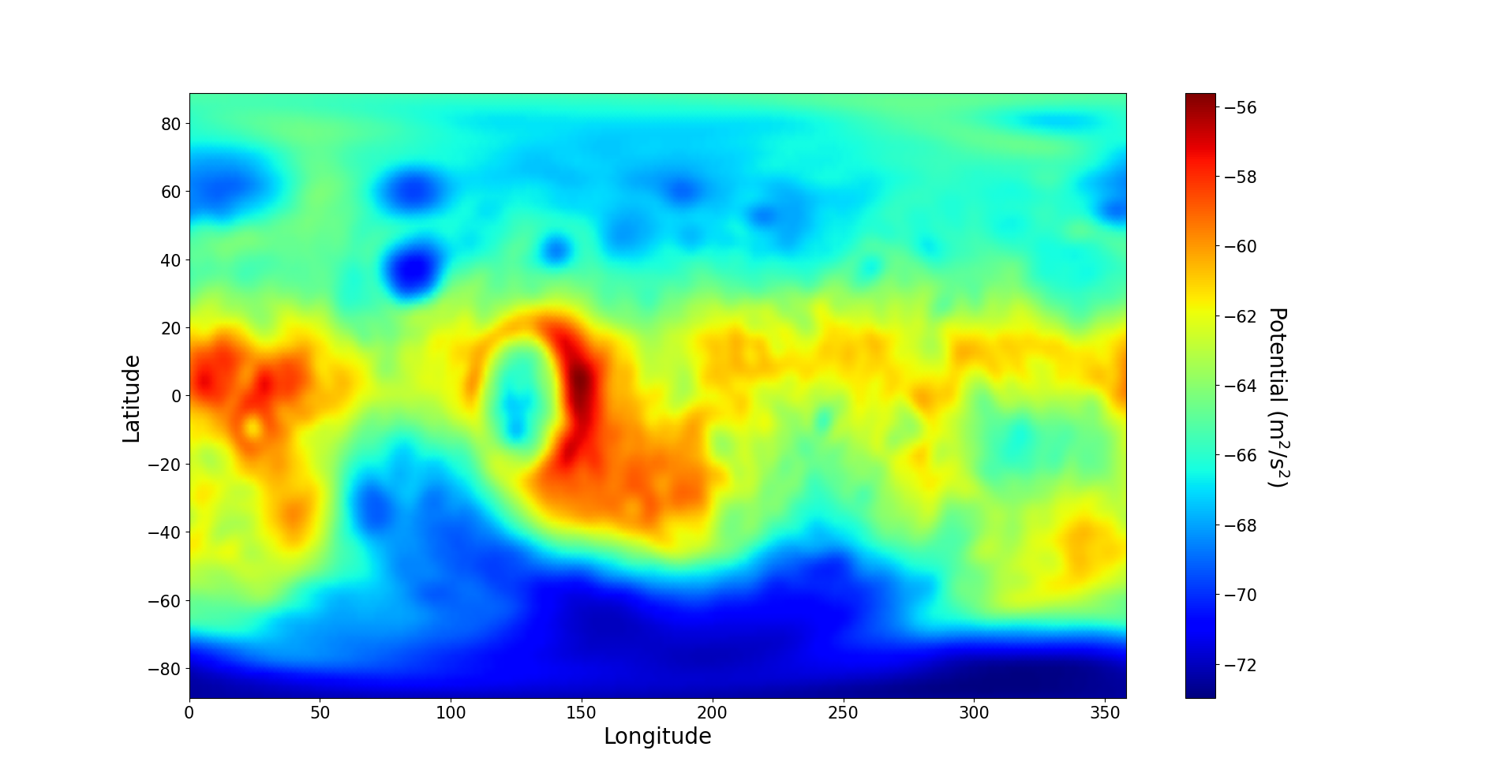}
    \caption{Surface potential (m$^2$s$^{-2}$) of Phobos calculated up to degree $l=100$, for a constant density model. The radial spectral element method was of order $5$ and there were 10 internal layers. The relative residual error in the iterative scheme was $10^{-12}$. }
    \label{fig:phobospotential}
\end{figure}

The potential calculated at the surface for a homogeneous density distribution is shown in~\figref{fig:phobospotential}. With a relative error of $10^{-12}$ in the iterative solution for the potential, the calculation time takes $\sim$47 s, run on a single core of an AMD EPYC 9334. Memory requirements are not significant due to the use of the matrix-free method. Solution times can be reduced with a commensurate increase in relative error.

\begin{figure}
    \includegraphics[trim={2cm 0.5cm 4.6cm .5cm}, clip,width=\textwidth]{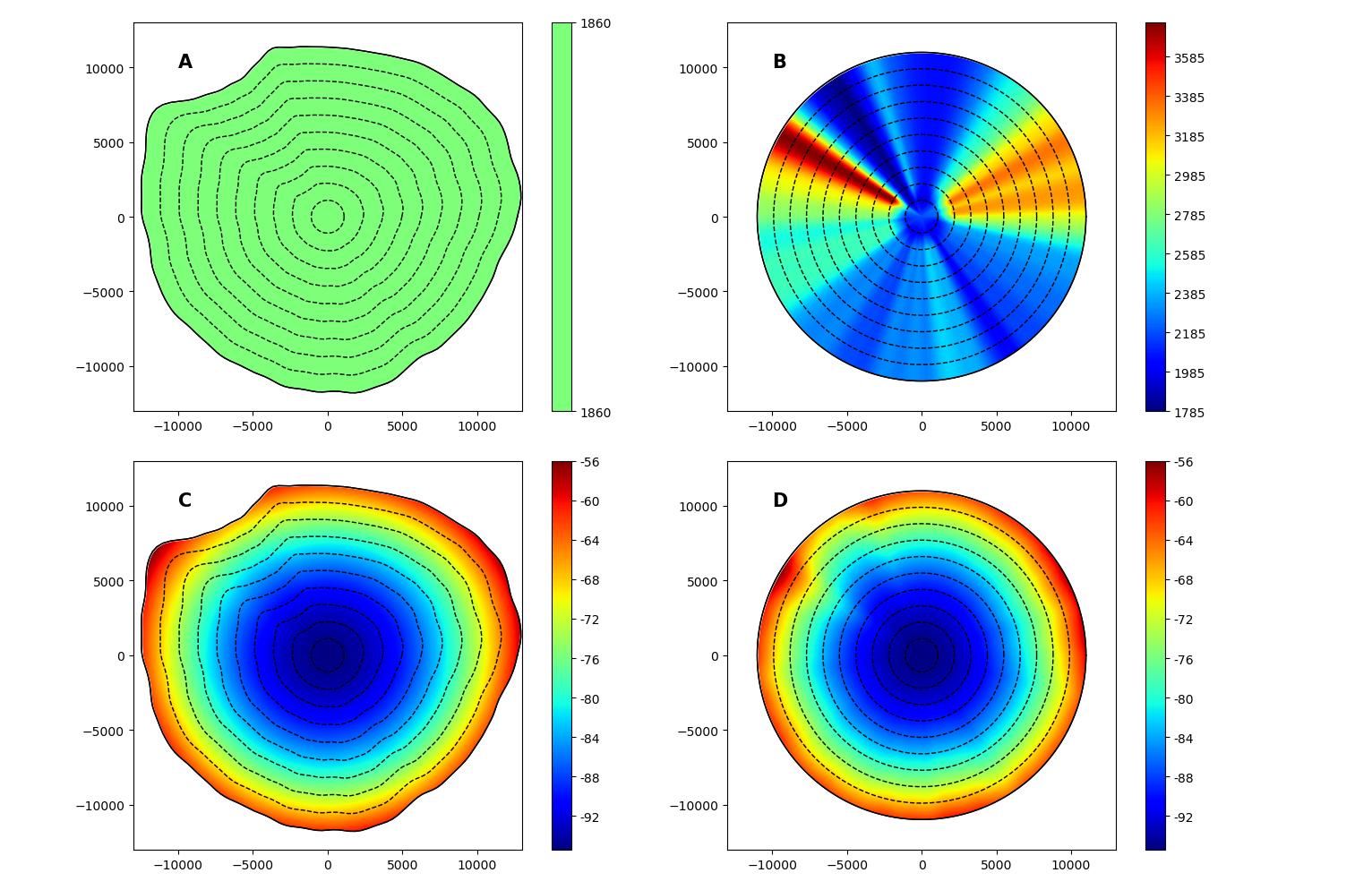}
     \caption{Comparison of the physical and referential densities and potentials in a slice through the prime meridian, for a homogeneous density Phobos. The potential was calculated correct up to $l=100$, with a fifth order spectral element method used radially and ten internal elements. The solid black lines indicate element boundaries in the computation and have no physical significance. (a) The physical density. (b) The referential density. (c) The physical potential. (d) The referential potential.}
    \label{fig:phobosphysicalvsreferential}
\end{figure}

It is informative to look at the physical and referential quantities in cross-sections through the planet. In particular, in~\figref{fig:phobosphysicalvsreferential} we compare the density and potential in both physical and referential frames. The radial nature of the mapping is made clear here. In particular if one compares (a) and (b) in~\figref{fig:phobosphysicalvsreferential}, the topographic variations are reflected in the referential density along radial lines. 

\begin{figure}
    \includegraphics[trim={2cm 0.5cm 4.6cm .5cm}, clip,width=\textwidth]{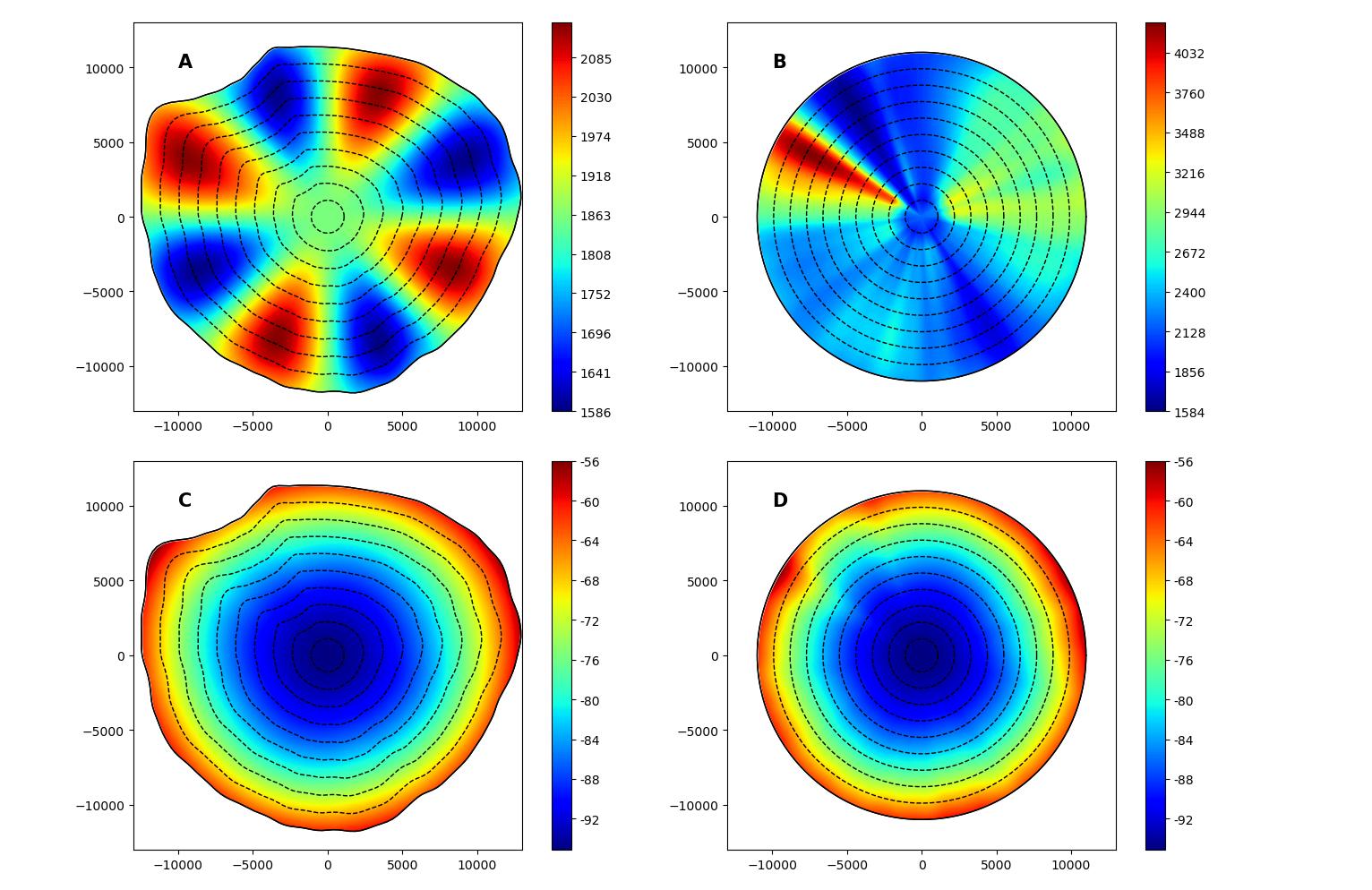}
     \caption{Comparison of the physical and referential densities and potentials in a slice through the prime meridian, for a heterogeneous density Phobos. The potential was calculated correct up to $l=100$, with a fifth order spectral element method used radially and ten internal elements. The solid black lines indicate element boundaries in the computation and have no physical significance. (a) The physical density. (b) The referential density. (c) The physical potential. (d) The referential potential.}
    \label{fig:phobosphysicalvsreferentialhetero}
\end{figure}

In~\figref{fig:phobosphysicalvsreferentialhetero} we perform the same calculation and look at the physical and referential quantities for a heterogeneous density model. The radial nature of the mapping can again be seen. It is striking that the main variation in the referential density still relates to the volume mapping rather than the (albeit relatively small) physical density variations.

\begin{figure}
    \includegraphics[width=\columnwidth,trim={4.0cm 1cm 4.1cm 2cm},clip]{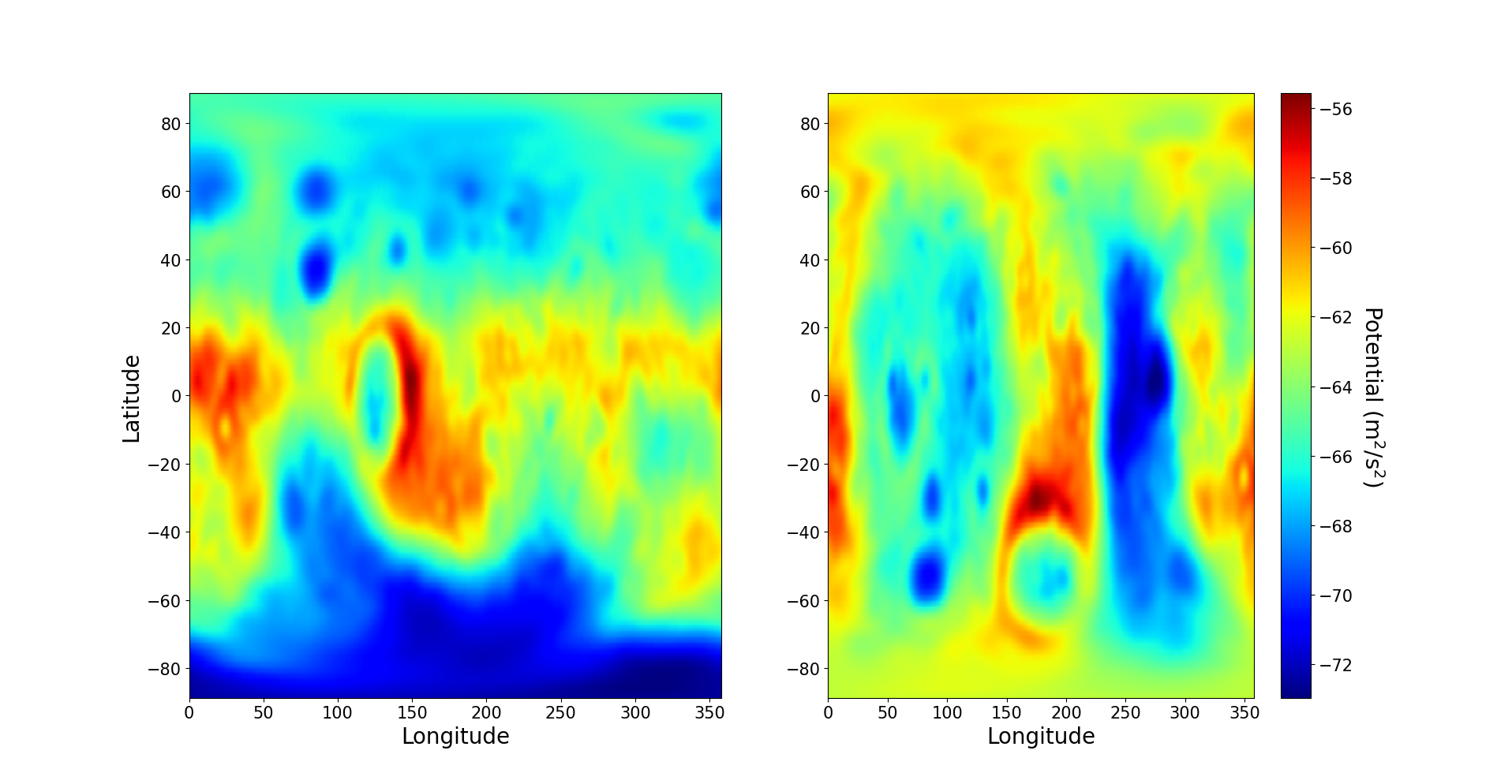}
    \caption{Depiction of the rotation functionality provided within the package. In this example, the gravitational potential on the surface of Phobos has firstly been calculated. Then, a slice along the prime meridian has been rotated to the equator, i.e. the slice $\phi = 0$ has been rotated so that it is now at $\theta = \pi/2$.}
    \label{fig:phobosrotated}
\end{figure}

Finally, it should be noted that any arbitrary cross-section can be obtained using rotations to the equator. This is implemented using the standard Euler angle method, which relates the relative orientation before and after rotation by three angles. We consider only passive transformations, whereby the axes used to describe the object are rotated, rather than rotating the object in physical space. To demonstrate this method, an example is shown in \figref{fig:phobosrotated}, whereby the prime meridian is rotated to the equator.

\subsubsection{Error tolerance}
The  linear system is solved using the pre-conditioned conjugate gradient method. The error estimated by the solver is based on the residual, in other words $||\mbf{A}\bs{\zeta}_{i} - \mbf{f}||/||\mbf{f}||$, where $\bs{\zeta}_{i}$ is the $i$-th approximation to the solution and the standard vector norm is being employed. This places the bound on the relative error in the solution vector as 
\begin{equation}
    \frac{||\bs{\zeta}- \mbf{A}^{-1}\mbf{f}||}{||\mbf{A}^{-1}\mbf{f}||} \leq \kappa(\mbf{A})  \frac{||\mbf{A}\bs{\zeta}-\mbf{f}||}{||\mbf{f}||},
\end{equation}
where $\kappa(\mbf{A})$ is the condition number of $\mbf{A}$ \cite{kincaid2009numerical}. As $\mbf{A}$ is Hermitian, it follows that $\kappa(\mbf{A}) = \lambda_{\text{max}}/\lambda_{\text{min}}$, where $\lambda_{\text{max}}$ is the maximum/minimum magnitude eigenvalue. Now, as we employ pre-conditioning, it is the eigenvalues of $\mbf{M}\mbf{A}$ which matter, i.e. the pre-conditioned matrix. Pre-conditioning, at least ideally, ensures the eigenvalues of $\mbf{M}\mbf{A}$ are all close to one. Consequently, it is reasonable to assume that, at least roughly,
\begin{equation}
    \frac{||\bs{\zeta} - \mbf{A}^{-1}\mbf{f}||}{||\mbf{A}^{-1}\mbf{f}||} \sim \frac{||\mbf{A}\bs{\zeta}-\mbf{f}||}{||\mbf{f}||}.
\end{equation}

The implication for our calculations is that the relative error of the solution, in a normed sense, is similar in magnitude to the relative error of the residual, which we can control. It is not possible to make a definitive statement about the pointwise error. However, calculations appear to show that the pointwise error is often lower than the residual error, particularly at large error tolerances. It is notable that there is a computational acceleration of a factor of two associated with changing the accuracy from $10^{-12}$ to $10^{-6}$. It is foreseeable that in some applications sacrificing accuracy for computational speed may be reasonable.

\section{Sensitivity analysis} \label{sec:sensitivity}

\subsection{Forward sensitivity theory} \label{sec:forsense}
The referential formulation enables the spherical harmonic expansion, as discussed above. It also implies that perturbations to the planetary shape (excepting those that create or intermix layers), can be accommodated without changing the geometry of the reference body. By this we mean that the number of layers, and the lower and upper radii of each layer within the reference body, are unaltered. To this end, all perturbations which we consider leave the geometric properties of the referential planet the same. 

The two parameters which can be perturbed are the referential density $\rho$ and the mapping $\bs{\xi}$. There are three different ``flavours'' of perturbation:
\begin{enumerate}
    \item Changing the referential density only. 
    \item Changing the mapping only, i.e. $\bs{\xi} \rightarrow \bs{\xi} + \delta \bs{\xi}$. This is, physically speaking, an elastic deformation of the planet\textemdash it is the deformation a planet will be subjected to when it is affected by the propagation of an elastic disturbance. It is noteworthy that the physical density does change, as one expects. 
    \item The final type of perturbation is a change to both the mapping and referential density. This is clearly the most general perturbation one can have. In the context of inverse problems, it is the case that naturally arises, if one is investigating the shape and density of a planet. 
\end{enumerate}
We now consider the effect of the three different types of perturbations on the potential. All calculations are performed up to first order\textemdash the effect of changing both the density and mapping can be found by adding up the contributions of the two effects.

\subsubsection{Changing the referential density}
The equation for the potential is, in operator form, $\mathcal{A}(\zeta,\chi)= f(\rho,\chi)$. Perturbing $\rho$ does not change the form of the operator $\mathcal{A}(\cdot,\cdot)$. It does, however, add a new force term given by
\begin{equation}
    \delta f = -4\pi G \int_{\tilde{{\mm}}} \delta \rho \overline{\chi} \dd^3 \bx.
\end{equation}
Due to linearity, the corresponding change in the potential is found by solving $\mbf{A} \delta\bs{\zeta} = \delta \mbf{f}$, where $\delta\bs{\zeta}$ is the change in the spherical harmonic components of the potential and $\delta \mbf{f}$ is the change in the force vector. In the computational framework, finding this perturbation is straightforward and only requires finding the force term corresponding to the density perturbation.

\subsubsection{Changing the mapping $\bs{\xi}$}
We now consider a change in the mapping, i.e. $\bs{\xi} \rightarrow \bs{\xi} + \delta \bs{\xi}$. The weak form for the problem, before the perturbation to the planetary shape, is given by~\eqref{eqn:weakpot0}. Under the perturbation $\bs{\xi} \rightarrow \bs{\xi} + \delta \bs{\xi}$, we wish to find an equation for the potential perturbation $\delta\zeta$ where $\zeta \rightarrow \zeta + \delta\zeta$. The only place in the equation where the mapping is present is in the term $\mbf{a}$. Given the definition of $\mbf{F}$ we have $\delta \mbf{F} = (\nabla \delta \bs{\xi})^T$. Using the binomial expansion we have, to first order, $\delta (\mbf{F}^{-1}) = -\mbf{F}^{-1} \delta \mbf{F} \mbf{F}^{-1}$. Similarly, for $J = \det \mbf{F}$, we have $\delta J = J \text{tr} (\mbf{F}^{-1} \delta \mbf{F})$. This implies that
\begin{equation}
    \delta \mbf{a} = J \mbf{F}^{-1} \left[ J^{-1} \delta J - \delta \mbf{F} \mbf{F}^{-1}  -  \mbf{F}^{-T} \delta \mbf{F}^T\right] \mbf{F}^{-T}.
\end{equation}
Ignoring the terms of $\mathcal{O}(\delta^2)$ in the equation and using the unperturbed equation~\eqref{eqn:weakpot0} we find the equation for the potential perturbation is 
\begin{equation}
    \int_{\mb} \left<\mbf{a} \nabla \delta \zeta, \nabla\bar{\chi} \right> \dd^3 \mathbf{x} + \sum_{lm} (l + 1) b \delta \zeta_{lm}(b) \bar{\chi}_{lm}(b) = - \int_{\tilde{\mm}} \left<\delta \mbf{a} \nabla \zeta, \nabla\bar{\chi} \right> \dd^3 \bx. \label{eqn:pertpot}
\end{equation}
It is important to note that the only change from \eqref{eqn:weakpot0} to \eqref{eqn:pertpot} is a change of force term. The left hand side remains the same, which implies that if we have solved for the potential in the unperturbed planet, any decompositions made already (such as the Cholesky decomposition of the pre-conditioner), can be reused.

\subsubsection{The overall change in force term} \label{sec:fullpertforce}
We now consider the change in the force term if both the referential density and the mapping are changed. The linearity of the problem implies that the contributions to the force term can be added, giving the final equation for the potential perturbation $\delta \zeta$ as 
\begin{equation}
    \int_{\mb} \left<\mbf{a} \nabla \delta \zeta, \nabla\bar{\chi} \right> \dd^3 \mathbf{x} + \sum_{lm} (l + 1) b \delta \zeta_{lm}(b) \bar{\chi}_{lm}(b) = -4\pi G \int_{\tilde{{\mm}}} \delta \rho \overline{\chi} \dd^3 \bx - \int_{\tilde{\mm}} \left<\delta a \nabla \zeta, \nabla\bar{\chi} \right> \dd^3 \bx. \label{eqn:pertpotfinal}
\end{equation}

\subsection{Phobos}
In addition to the application in the context of inverse problems, another application of the perturbation method is in lowering the cost of calculations. Many planets, notably the Earth, are close to spherical\textemdash the Earth's ellipticity is on the order of $1/300$. Consequently, for some applications, it is foreseeable that approximating the potential using first-order perturbation theory may be reasonable if it comes with a commensurate decrease in the cost.

\begin{figure}
    \includegraphics[width=\columnwidth,trim={4.3cm 1.3cm 4cm 2cm},clip]{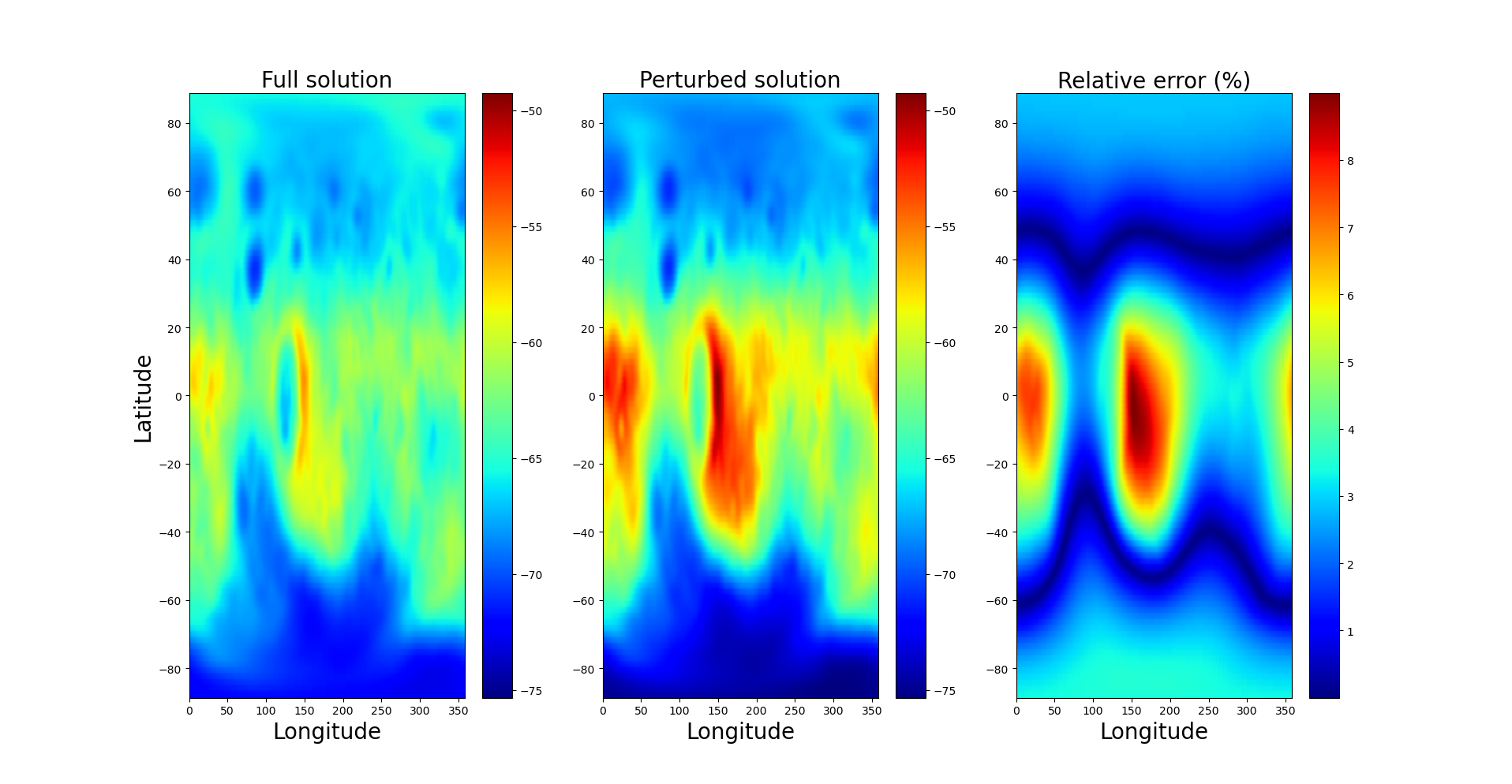}
        \caption{Surface potential calculations using forward perturbation theory for Phobos. The exact calculation is performed to a relative error of $10^{-12}$ and includes all topography, with $l_{\text{max}} = 100$ for the computation, and using order 5 radial spectral elements. The base cases for the perturbation calculations is performed including topography up to $l_{\text{max}} = 0$ and the iterative solver is applied with a relative error of $10^{-3}$.}
        \label{fig:phobosperturbationa}
\end{figure}

\begin{figure}
    \includegraphics[width=\columnwidth,trim={4.3cm 1.3cm 4cm 2cm},clip]{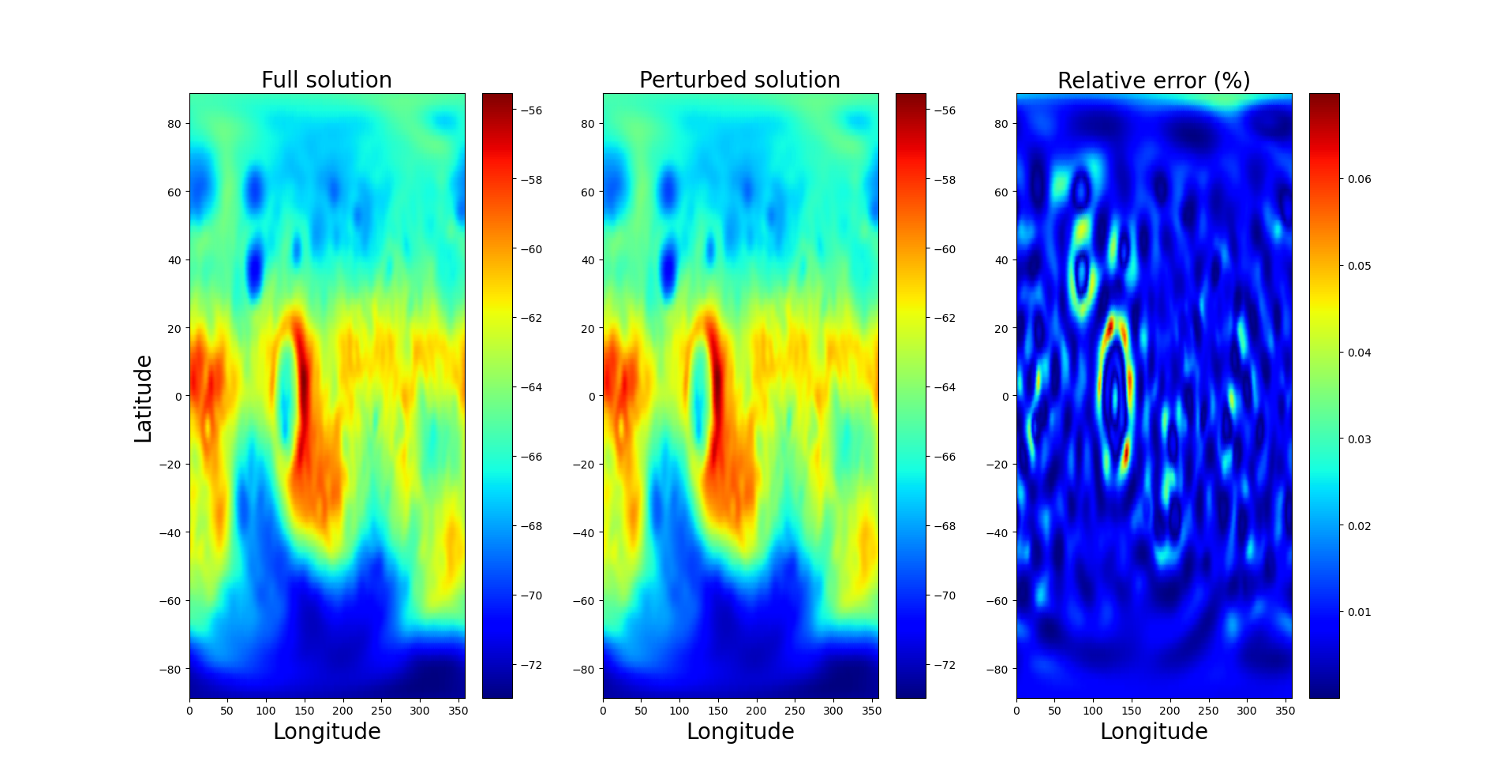}
       \caption{Surface potential calculations using forward perturbation theory for Phobos. The exact calculation is performed to a relative error of $10^{-12}$ and includes all topography, with $l_{\text{max}} = 100$ for the computation, and using order 5 radial spectral elements. The base cases for the perturbation calculations is performed including topography up to $l_{\text{max}} = 15$ and the iterative solver is applied with a relative error of $10^{-3}$.}
        \label{fig:phobosperturbationc}
\end{figure}

To demonstrate the utility of the perturbation method when used in this context, we reconsider the case of Phobos. The radial displacement $h(\theta,\varphi)$ can be split into two parts $h(\theta,\varphi) = h_0(\theta,\varphi) + \delta h(\theta,\varphi)$. The maximum degree spherical harmonic in $h_0$ is chosen to be $l_{\text{approx}}$, which implies that
\begin{equation}
    h_0(\theta,\varphi) = \frac{r}{r_{\text{av}}}\sum_{l = 0}^{l_{\text{approx}}} \sum_{m = -l}^{l} h_{lm} Y_{lm}, \hspace{2cm}
    \delta h(\theta,\varphi) = \frac{r}{r_{\text{av}}}\sum_{l = l_{\text{approx}} + 1}^{l_{\text{max}}} \sum_{m = -l}^{l} h_{lm} Y_{lm}.
\end{equation}
The zero-th order referential model of Phobos is defined using $h_0(\theta,\varphi)$. The gravitational potential $\phi_0$ is calculated in this model. The perturbation to the potential $\delta \phi$ associated with $\delta h(\theta,\varphi)$ can then be calculated. To determine the errors associated with the perturbation method the exact potential is also calculated. This procedure is demonstrated in~\figref{fig:phobosperturbationa} and~\figref{fig:phobosperturbationc} for different values of $l_{\text{max}}$. 

These examples were chosen with $l_{\text{max}} = 0$ and $15$ for~\figref{fig:phobosperturbationa}, and \figref{fig:phobosperturbationc} respectively. The error associated with the use of the perturbation method in \figref{fig:phobosperturbationa} is on the order of a few percent. Although this may seem large, the unperturbed model is spherical, and perturbations to the radius are on the order of several tens of percent. For an unperturbed model including all topography up to degree five (not shown), the error is on the order of a few tenths of a percent. When including all topography to degree fifteen, the error is on the order of a few hundredths of a percent. This behaviour aligns with expectation; as the initial model approaches the exact model, the perturbation to the potential, and the overall error, decreases commensurately.

The computational cost associated with the calculations, when compared with that of the exact method, is, at first, surprising. In the case with a spherical initial model, the cost was around one order of magnitude lower. As $l_{\text{max}}$ for the initial model increased, however, this effect rapidly diminished. In fact, for the calculations with $l_{\text{max}} = 5,15$, the computational times were very similar for the exact and perturbation methods. The inclusion of topography in the unperturbed model makes the preconditioning less effective. This means that the convergence is faster than in the exact method, but may not be significantly so. In the case of Phobos, there is also a relatively slow decrease in the power with increasing degree in the spherical harmonic expansion. There is, however, a cost saving associated with the perturbation method, arising from the error tolerance in the perturbation calculation. To address this we need to consider the solution as being $\bs{\zeta} = \bs{\zeta}_0 + \delta \bs{\zeta}$. To first order in $\delta$ we have
\begin{equation}
    ||\bs{\zeta}|| \approx ||\bs{\zeta}_0|| \left[1 + \hat{\bs{\zeta}}_0 \cdot \frac{\delta \bs{\zeta}}{||\bs{\zeta}_0||}\right].
\end{equation}
The perturbation solution constrains the error in the perturbation, i.e. $\frac{||e(\delta\bs{\zeta})||}{||\delta \bs{\zeta}||}$, where $e(\delta(\bs{\zeta}))$ is the error in $\delta\bs{\zeta}$. As the perturbation approach is only reasonable when $||\delta \bs{\zeta}|| \ll || \bs{\zeta}_0||$, it implies that the relative error in ${\bs{\zeta}}$ arising from the error in $\delta{\bs{\zeta}}$ is
\begin{equation}
    \frac{e(\delta \bs{\zeta})}{||\bs{\zeta}_0||} = \frac{||e(\delta\bs{\zeta})||}{||\delta \bs{\zeta}||} \cdot \frac{||\delta \bs{\zeta}||}{|| \bs{\zeta}_0||} \ll \frac{||e(\delta\bs{\zeta})||}{||\delta \bs{\zeta}||}.
\end{equation}
Consequently, the error in $\delta{\bs{\zeta}}$ associated with errors in the solution of the perturbation problem is much smaller than the error associated with using the perturbation method itself. The error tolerance used in the perturbation solution can, therefore, be relaxed. If used appropriately this can result in computational acceleration.

\section{Adjoint problem} \label{sec:adjoint}

\subsection{Sensitivity kernels for the gravity problem}
We use the Lagrangian method for PDE constrained optimization (e.g.~\citeA{troltzsch2010control})\textemdash whereby variational principles are used to give a systematic method to determine the sensitivity kernels. One finds, firstly, the adjoint equation, whose solution gives the sensitivity kernels we seek.

\subsubsection{The adjoint equation}
We consider a real-valued functional 
\begin{equation}
    Q(\zeta) \equiv \hat{Q}(\rho,\bs{\xi},\mbf{F}),
\end{equation}
where $Q(\zeta)$ is a function of $\zeta$ and $\hat{Q}(\rho,\bs{\xi},\mbf{F}) =  Q[\hz(\rho,\bs{\xi},\mbf{F})]$ is termed the reduced objective functional. The perturbation in $\hat{Q}(\zeta)$ associated with a change in density or the mapping can be written as
\begin{equation}
    \delta \hat{Q}(\zeta) = \int_{\tilde{\mm}} \left[K_{\rho} \delta \rho + \left<\mbf{K}_{\xi}, \delta \bs{\xi}\right> + \left<\mbf{K}_{F},\delta \mbf{F}\right>\right] \dd^3 \bx, \label{eqn:functionalpert}
\end{equation}
which defines the sensitivity kernels $K_{\rho} = (D_{\rho}f)^\dagger DQ(\zeta)$, $\mbf{K}_{\xi} = (D_{\bs{\xi}}f)^\dagger DQ(\zeta)$ and $\mbf{K}_{F}=(D_{\mbf{F}}f)^\dagger DQ(\zeta)$. To derive the adjoint equation, we define a Lagrangian 
\begin{align}
    \mathcal{L} &= Q(\zeta) +  \int_{\mb} \left<\mbf{a} \nabla \zeta, \nabla{\chi} \right> \dd^3 \mathbf{x} + \sum_{lm} (l + 1) b \zeta_{lm}(b) \chi_{lm}(b)  + 4 \pi G \int_{\tilde{\mm}} \rho\chi \dd^3 \mathbf{x}.
\end{align}
In this expression, $Q$ is the functional whose sensitivity kernel we seek to obtain. The remaining terms are familiar as the weak form of the forward problem. The potential $\zeta$ is associated with an adjoint state variable $\chi$. If $\mathcal{L}$ is stationary with respect to variations in $\chi$ then $\zeta$ solves the forward problem\textemdash it gives the potential associated with $(\rho,\bs{\xi},\mathbf{F})$. If $\mathcal{L}$ is then required to be stationary with respect to $\zeta$ then the Lagrange multiplier theorem states that the functional derivatives of $\mathcal{L}$ and $Q$ with respect to $(\rho,\bs{\xi},\mbf{F})$ are the same~\cite{troltzsch2010control}. The variation of $\mathcal{L}$ with respect to $\zeta$ gives
\begin{align}
    \delta \mathcal{L}|_{\zeta} &= \left<D_{\zeta}Q,\delta \zeta\right> +  \int_{\mb} \left<\mbf{a} \nabla \delta \zeta, \nabla{\chi} \right> \dd^3 \mathbf{x} + \sum_{lm} (l + 1) b \delta \zeta_{lm}(b) \chi_{lm}(b),
\end{align}
where we use the notation $\delta \mathcal{L}|_{\zeta}$ to indicate the perturbation in $\mathcal{L}$ associated with a perturbation in $\zeta$, whilst all other parameters are held constant. Relabelling $\delta \zeta \rightarrow \bar{\eta}$ for ease of notation, using the Hermiticity of $\mbf{a}$~(MA19), and setting the variation to zero we find the adjoint equation
\begin{align}
    \int_{\mb} \left<\mbf{a} \nabla \chi, \nabla\bar{\eta} \right> \dd^3 \mathbf{x} + \sum_{lm} (l + 1) b \chi_{lm}(b) \bar{\eta}_{lm}(b) = -\left<D_{\zeta}Q,\delta \zeta\right> . \label{eqn:adjoint}
\end{align}
The importance of the adjoint equation will be made clear in the subsequent subsections. It is interesting to note that the adjoint equation for $\chi$ is also a Poisson equation. Computationally, it is only the force term that we need to determine, as the machinery to solve the forward problem can be used for the left hand side.

\subsubsection{The sensitivity kernels}
We seek to find the sensitivity kernels for density $\rho$, the mapping $\bs{\xi}$ and the deformation gradient $\mbf{F}$. To determine the sensitivity kernels, we perturb the appropriate parameter in the Lagrangian and find the term that is multiplied by the perturbation:
\begin{enumerate}
    \item The only place where density appears is in the final term in $\ml$. Consequently, we have     
    \begin{align}         
        \delta \ml|_{\rho} &= 4\pi G \int_{\tmm} \chi \delta \rho \dd^3 \bx.     
    \end{align} 
    The sensitivity kernel for density is then given by
    \begin{equation}
        K_{\rho} = 4 \pi G \chi,
    \end{equation}
    i.e. it is directly proportional to the adjoint variable.
    \item The sensitivity kernel for the mapping $\bs{\xi}$ is trivially given by $\mbf{K}_{\bs{\xi}} = 0$, as the Lagrangian does not depend explicitly on $\bs{\xi}$\textemdash the objective functional does not have dependence on $\bs{\xi}$.
    \item The sensitivity kernel for the deformation gradient $\mbf{F}$ requires more calculation. In particular, the objective functional $Q$ may have some dependence upon $\mbf{F}$. However, the observations that we will use are often on a surface enclosing the planet, e.g. satellite data. The mapping and its perturbation can be chosen to be zero on this surface which implies that one can define the problem such that there is no direct sensitivity to $\mbf{F}$. Thus
    \begin{equation}
        \delta \ml|_{\mbf{F}} = \int_{\mb} \left<\delta \mbf{a} \nabla \zeta, \nabla \chi\right> \dd^3 \bx. \label{eqn:dl}
    \end{equation}
    It was shown earlier that $\delta \mbf{a} = \mbf{a} \text{Tr}[\mbf{F}^{-1}\delta \mbf{F}] - \mbf{F}^{-1} \delta \mbf{F} \mbf{a} - \mbf{a} \delta \mbf{F}^T \mbf{F}^{-T}$ (using $\text{Tr}$ for the trace of a tensor). Using this identity with~\eqnref{eqn:dl}
    \begin{align}
        \left<\mbf{K}_{F},\delta \mbf{F}\right> &= \left<\mbf{a}\text{Tr}[\mbf{F}^{-1}\delta \mbf{F}] \nabla \zeta, \nabla \chi\right> - \left<\mbf{F}^{-1}\delta \mbf{F} \mbf{a} \nabla \zeta, \nabla \chi\right> - \left<\mbf{a}\delta \mbf{F}^T \mbf{F}^{-T} \nabla \zeta, \nabla \chi\right>, \nonumber\\
        &= \left<\mbf{a} \nabla \zeta, \nabla \chi\right> \left<\mbf{F}^{-T},\delta \mbf{F}\right> - \left<\delta \mbf{F} \mbf{a} \nabla \zeta, \mbf{F}^{-T}\nabla \chi\right> - \left<\mbf{F}^{-T} \nabla \zeta, \delta \mbf{F} \mbf{a}^T \nabla \chi\right>.
    \end{align}
    Using the symmetry of $\mbf{a}$ and writing $\otimes$ for tensor products we have
    \begin{equation}
        \mbf{K}_{F} = \left<\mbf{a} \nabla \zeta, \nabla \chi\right>  \mbf{F}^{-T} - (\mbf{F}^{-T} \nabla\chi) \otimes (\mbf{a} \nabla \zeta) - (\mbf{F}^{-T} \nabla \zeta) \otimes (\mbf{a} \nabla \chi).
    \end{equation}
    It is notable that the sensitivity kernel for deformation gradient is dependent upon both the potential and adjoint potential.
\end{enumerate}

\subsubsection{Summary}
The procedure for calculating the sensitivity kernels can be summarised as:
\begin{enumerate}
    \item Solve the Poisson equation~\eqref{eqn:weakpot0} for potential $\zeta$.
    \item Solve the adjoint equation~\eqref{eqn:adjoint} for adjoint variable $\chi$. As the adjoint equation is also a Poisson equation one only needs to implement a function to find the adjoint force term.
    \item Use the solutions to the Poisson and adjoint problem to find the sensitivity kernels $K_{\rho}$ and $\mbf{K}_F$.
\end{enumerate}
It should be emphasised that the form of the equations is independent of the choice of the objective functional $Q(\zeta)$, and its effect is confined to the adjoint force term. The analytical form of the adjoint force term will vary (and the numerical techniques required to evaluate it), depending upon the definition of $Q(\zeta)$.

\subsection{Phobos}
We determine some sensitivity kernels for referential density in the case of Phobos. We choose to look at a spherical harmonic component of the potential on a ball surrounding the moon. The functional whose sensitivity kernel we wish to find is 
\begin{equation}
    Q(\zeta) = b^{-2}\int_{\partial \mb} \zeta(b,\theta,\varphi) \bar{Y}_{lm} \dd S.
\end{equation}
In the case of a geometrically spherical planet the $Y_{lm}$ component of the potential on a surface can be written in terms of a radial integral of the same spherical harmonic component of density, in particular
\begin{equation}
    Q(\zeta) = -\frac{4\pi G}{2l + 1} b^{-(l + 1)} \int_{0}^r r'^{2 + l} \rho_{lm}(r') \dd r' ,
\end{equation}
where details are presented in Appendix B. However, Phobos is highly aspherical, which suggests that the referential sensitivity kernels for a single spherical harmonic should have power in multiple spherical harmonics.

\begin{figure}
    \includegraphics[width=\columnwidth, trim = {16cm 2cm 7cm 2cm}, clip]{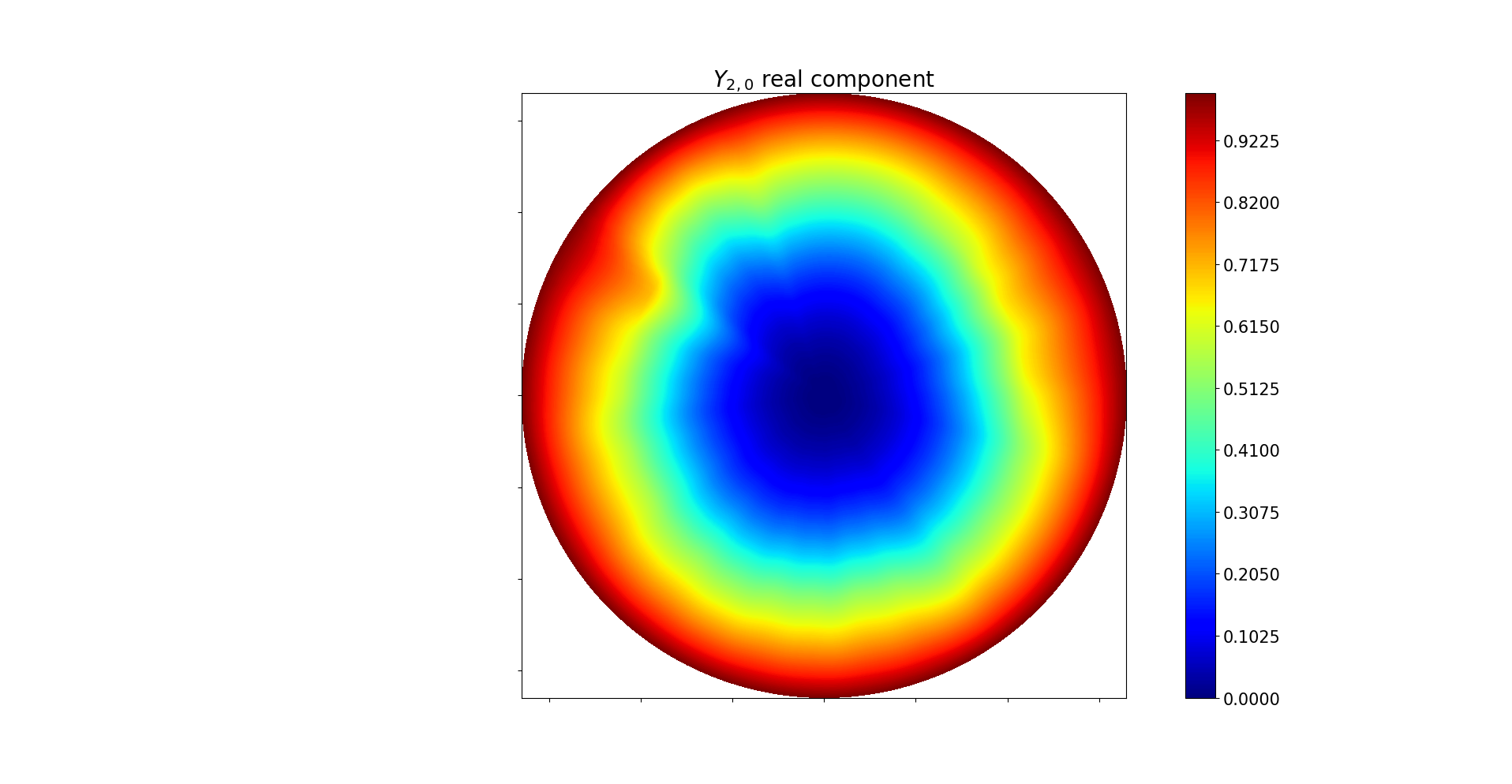}
    \caption{Real component of (non-dimensionalised) sensitivity kernel for the $Y_{20}$ component of the potential on a ball surrounding Phobos at a radius of 16.5km. The cross-section through the planet is taken along the equator.}
    \label{fig:sensitivityphobos}
\end{figure}

A calculation of the sensitivity kernel for the $Y_{20}$ component of potential, on a sphere of radius $16.5$ km surrounding Phobos is shown in~\figref{fig:sensitivityphobos}. The cross-section of the sensitivity kernel is taken through the equator. The $Y_{20}$ spherical harmonic has no azimuthal dependence and consequently in a geometrically spherical planet it will be axially symmetric about the z-axis. However, as one can see in~\figref{fig:sensitivityphobos}, the sensitivity kernel is not axially symmetric; the geometry is mapped into the sensitivity kernel. This is a natural consequence of the particle relabelling transformation. 

To use the sensitivity kernel to determine the change in the functional $Q$ for arbitrary perturbations to the referential density, one must simply integrate the density perturbation against the sensitivity kernel throughout the referential domain. Although this example, of the spherical harmonic component, could probably be addressed without resorting to the general approach detailed here, the advantage of a referential approach is nonetheless clear, especially in the geometric simplicity of the domain.

\subsection{A general approach to referential perturbation expressions}
The perturbation to an objective functional given in~\eqref{eqn:functionalpert} involves an integral over the planetary body with three separate sensitivity kernels for density, the mapping and the deformation gradient. In non-referential approaches, perturbations can typically be decomposed into two parts, an integral over the body and a surface integral over its boundary, see for instance equation (3.98) of~\citeA{Dahlen_Tromp_1998}. The advantage of expressing the perturbation in this form is both simplicity and physical intuition. It is the goal of this subsection to demonstrate how this may be done quite generally within a referential formulation. Ultimately we obtain a ``decomposition'' into two integrals, and, as an intermediate step, a general identity which must be satisfied by a set of sensitivity kernels, for any objective functional in the referential formulation.

\subsubsection{The referential sensitivity kernel identity}
We use $Q$ to indicate the objective functional, and do not place any restrictions on the functional except that it is consistently defined in physical space, i.e. its value is independent of the referential body and mapping chosen. This implies that if the referential density and the mapping are simultaneously perturbed, such that the physical planet remains unaltered, $Q$ must also be unaltered. To effect this type of perturbation, we consider the equation for the referential density and the effect of a perturbation in the mapping. In other words we define $\bs{\xi} \rightarrow \bs{\xi} + \delta \bs{\xi}$, and choose a $\delta \rho$ such that $\varrho$ remains unperturbed. The perturbation $\delta \bs{\xi}$ on the boundaries $\delta \tmm$ must be zero. We have then
\begin{align}
    (\rho + \delta \rho)(x) & = (J + \delta J)(x) [\varrho \circ (\xi + \delta \bs{\xi})](x), \nonumber\\
    &= J(x) [\varrho \circ \bs{\xi}] (x) + \delta J(x) [\varrho \circ \bs{\xi}](x) + J \delta \bs{\xi} \cdot \nabla (\varrho)({\bs{\xi}(x)}) + \mathcal{O}(\delta^2).
\end{align}
We recall that derivatives in the physical and referential space are linked by the transpose of the deformation gradient, i.e. for a function $\tilde{g}(x) = (g\circ \bs{\xi})(x)$, the gradient is given by $(\nabla \tilde{g})(x) = \mbf{F}^{T} (\nabla g)({\bs{\xi}(x)})$. This implies that 
\begin{equation}
    (\nabla \varrho)({\bs{\xi}(x)}) = \mbf{F}^{-T} \nabla(J^{-1}\rho ).
\end{equation}
Consequently, the perturbation in the referential density associated with no change to the physical planet is
\begin{equation}
    \delta \rho = \delta J J^{-1} \rho + J \delta \bs{\xi} \cdot \mbf{F}^{-T} \nabla(J^{-1} \rho).
\end{equation}
The Piola identity states that $\partial_i[J(F^{-1})_{ij}] =0$ (e.g.~\citeA{marsden1994mathematical}, theorem 7.20). Furthermore, $\delta J = J \text{tr}[\mbf{F}^{-1}\delta \mbf{F}] = J (F^{-1})_{ij} \partial_i \delta\xi_j$. Consequently, we can rewrite $\delta \rho$ as
\begin{align}
    \delta \rho &= [J^{-1}\rho]  [J (F^{-1})_{ij}] [\partial_i \delta\xi_j] + [J^{-1} \rho] \partial_i[J(F^{-1})_{ij}] [\delta \xi_j] + \partial_i[J^{-1} \rho] [J (F^{-1})_{ij}] [\delta \xi_{j}], \nonumber\\
    & = \partial_i [J^{-1} \rho J (F^{-1})_{ij} \delta \xi_j ], \nonumber\\
    &= \nabla \cdot [\rho \mbf{F}^{-1} \cdot \delta \bs{\xi}].
\end{align}
To proceed in determining the relationship between the sensitivity kernels we use Gauss' law on the integral involving the perturbation to the deformation gradient. This gives
\begin{align}
    \int_{\tmm} \left<\mbf{K}_{{F}}, \delta \mbf{F}\right> \dd^3 \bx & = \int_{\tmm} (\mbf{K}_{F})_{ij} \partial_j \delta \xi_i \dd^3 \bx, \nonumber\\
    &= \int_{\partial\tmm} \hat{n}_j  (\mbf{K}_F)_{ij} \delta \xi_i \dd S - \int_{\tmm} \delta \xi_i \partial_j (\mbf{K}_F)_{ij} \dd^3 \bx.
\end{align}
Now, the condition on $\delta \bs{\xi}$ is that it is identically zero on $\partial \tmm$. Consequently, we now have
\begin{equation}
    \delta Q = \int_{\tmm} K_{\rho} \delta \rho \dd^3 \bx + \int_{\tmm} \delta \xi_i [(\mbf{K}_{\xi})_i - \partial_j (\mbf{K}_F)_{ij}]\dd^3 \bx.
\end{equation}
Substituting the $\delta \rho$ that coincides with no physical change we have for the first term
\begin{align}
    \int_{\tmm} K_{\rho} \delta \rho \dd^3 \bx & = \int_{\tmm} K_{\rho} \nabla \cdot [\rho \mbf{F}^{-1} \delta \bs{\xi}] \dd^3 \bx, \nonumber\\
    &=\int_{\tmm}\nabla\cdot [K_{\rho}\rho \mbf{F}^{-1} \delta \bs{\xi}] \dd^3\bx - \int_{\tmm} (\nabla K_{\rho}) \cdot \rho \mbf{F}^{-1} \delta \bs{\xi} \dd^3 \bx, \nonumber\\
    &=\int_{\partial\tmm} \hat{\mbf{n}} \cdot K_{\rho}\rho \mbf{F}^{-1} \delta \bs{\xi} \dd S - \int_{\tmm} \delta \bs{\xi} \cdot \rho \mbf{F}^{-T} \nabla K_{\rho} \dd^3 \bx.
\end{align}
As the perturbation $\delta \bs{\xi}  = 0$ on $\partial \tmm$ we have
\begin{equation}
    \delta \hat{Q}=  \int_{\tmm} \delta \xi_i [(K_{\xi})_i - \partial_j (K_F)_{ij} - \rho (F^{-T})_{ij} \partial_j K_{\rho}]\dd^3 \bx.
\end{equation}
Finally, this must be zero for all admissible variations $\delta \bs{\xi}$ and we then have
\begin{equation}
    \mbf{K}_{\xi} = \rho \mbf{F}^{-T} \nabla K_{\rho} + \nabla \cdot \mbf{K}_F, \label{eqn:refsensident}
\end{equation}
where we have defined $\nabla \cdot T \equiv \partial_i T_{j_1...j_{n-1} i}$ for a tensor $T_{j_1 j_2 ... j_n}$. It is this expression that we term the referential sensitivity kernel identity. It must hold for the referential sensitivity kernels of any function that depends on the density, mapping and deformation gradient alone.

\subsubsection{The final perturbation expression}
We are now in the position to rewrite the perturbation expression~\eqref{eqn:functionalpert} in terms of a body and a surface integral, employing~\eqnref{eqn:refsensident}. The starting point is
\begin{equation}
    \delta \hat{Q} = \int_{\tmm} K_{\rho} \delta \rho \dd^3 \bx + \int_{\tmm} \left<\mbf{K}_{\xi}, \delta \bs{\xi}\right> \dd^3 \bx + \int_{\tmm} \left<\mbf{K}_F, \delta \mbf{F}\right> \dd^3 \bx. \label{eqn:perturbl}
\end{equation}
Substituting~\eqref{eqn:refsensident} into the second term we have
\begin{align}
    \int_{\tmm} \left<\mbf{K}_{\xi}, \delta \bs{\xi}\right> \dd^3 \bx & = \int_{\tmm}  \delta \bs{\xi} \cdot \rho \mbf{F}^{-T} \nabla K_{\rho}  \dd^3 \bx + \int_{\tmm}  \delta \bs{\xi} \cdot (\nabla \cdot \mbf{K}_F) \dd^3 \bx. \label{eqn:dxiperturb}
\end{align}
The first of the terms in~\eqref{eqn:dxiperturb} can be rewritten as
\begin{align}
    \int_{\tmm} \delta \bs{\xi} \cdot \rho \mbf{F}^{-T} \nabla K_{\rho}  \dd^3 \bx & = \int_{\tmm} \rho\delta \xi_i  (\mbf{F}^{-T})_{ij} \partial_j K_{\rho}  \dd^3 \bx, \nonumber\\
    & = \int_{\tmm} \partial_j[\rho \delta \xi_i (\mbf{F}^{-T})_{ij} K_{\rho}] \dd^3 \bx - \int_{\tmm} K_{\rho} \partial_j [\rho \delta \xi_i (\mbf{F}^{-T})_{ij} ] \dd^3 \bx, \nonumber\\
    & = \int_{\partial \tmm} K_{\rho}\rho \delta \bs{\xi} \cdot \mbf{F}^{-T} \cdot \hat{\mbf{n}} \dd S - \int_{\tmm} K_{\rho} \nabla\cdot [\rho \delta \bs{\xi} \cdot \mbf{F}^{-T}] \dd^3 \bx.
\end{align}
Finally, the third term in~\eqref{eqn:perturbl} can be rewritten using the same method as in the previous subsection and is
\begin{equation}
    \int_{\tmm} \left<\mbf{K}_F, \delta \mbf{F}\right> \dd^3 \bx = \int_{\partial \tmm} \delta \bs{\xi} \cdot \mbf{K}_F \cdot \hat{\mbf{n}} \dd S - \int_{\tmm} \delta \bs{\xi} \cdot (\nabla \cdot \mbf{K}_F) \dd^3 \bx.
\end{equation}
Combining all the terms we have the result
\begin{equation}
    \delta \hat{Q} = \int_{\tmm} K_{\rho} \left\{\delta \rho - \nabla\cdot \left[\rho \mbf{F}^{-1} \cdot \delta \bs{\xi} \right]\right\} \dd^3 \bx + \int_{\partial \tmm} \delta \bs{\xi} \cdot \left[\mbf{K}_F + \rho K_{\rho} \mbf{F}^{-T}\right] \cdot \hat{\mbf{n}} \dd S.
\end{equation}
Consequently, the perturbation has been reduced to a single integral over the body and a single integral over the boundaries. The presence of the divergence term within the first integral is associated with an effective advection term in the referential body, induced by the perturbation to the mapping; the sum of the density perturbation and the advective term is only zero if the density perturbation is such that the physical density remains constant. The second integral, i.e. the one over the boundaries, is associated with advection across the boundary induced by the perturbation. This term combines the effect of the deformation gradient and the effect of the mass movement on the boundary.

\section{Conclusions}

In this paper we have expanded on the work of~MA19. It has been shown that the gravitational potential associated with a perturbation to the shape of a body can be easily evaluated using the same forward modelling approach. Sensitivity kernels have been derived for near-arbitrary measurements of potential. In addition, a simplification of the expression for the perturbation to an arbitrary objective functional has been derived. The code base has been thoroughly benchmarked, using analytical results, integral results in spherical geometry and different mappings for physically identical planets (shown in Appendix A). As an example of the use of the code, the gravitational potential of Phobos has been calculated. The example of Phobos has also been used to demonstrate the utility of the perturbation method, with an observation made of the improvement in convergence as the initial model approaches the exact model. Finally, the sensitivity kernels for a spherical harmonic component of the potential have been derived for Phobos. 

We hope that this code can be useful for a wide range of researchers. In particular, however, we hope that those interested in gravitational potential calculations and measurements in small bodies will be able to employ it. The calculation of sensitivity kernels for functions of potential is, in theory, straightforward. Nonetheless, if one wants to find the sensitivity kernel, separate calculations are often required to be made. The inclusion of this within the code will hopefully simplify this procedure. Our hope is that this will enable simpler implementation of gradient-based optimisation and other techniques requiring the use of sensitivity kernels.

\appendix


\section{Modelling benchmarks and calculations}
The benchmarking procedure was broken into two separate parts: firstly checking the code with no mapping and secondly checking that the inclusion of the mapping is correct. The first part is relatively simple as implementation of the spherical case in a ``hard-coded'' manner is straightforward (and forms the basis of the pre-conditioning). In addition, exact integrals of spherical harmonic components are trivial. In the examples folder in the package, the first few examples detail the benchmarks for geometrically spherical planets, in order of increasing complexity. The second stage of benchmarking is checking the implementation of the mapping. In this case there are several methods. One can have a mapping to a physical planet for which the calculation of the potential is straightforward, from an arbitrary referential planet. Secondly, there is the option of doing a ``self-benchmark'', where the potential for an arbitrary physical planet is calculated using two different referential planets and the results are compared.

In the following subsections we detail two benchmarks which we believe best illustrate the ideas discussed above. In the interest of demonstrating the efficacy of the mapping, both have non-identity mappings. For further examples the reader is referred to the package itself.

\subsection{Forward modelling}
\subsubsection{Radial dilation}
The first example chosen is that of pure radial dilation, i.e. where $\bs{\xi} = (1 + h) \bx$. The reference planet is chosen to be a homogeneous sphere of radius $a$ and constant density $\rho_0$. The physical planet is a homogeneous sphere of radius $(1 + h)a$ and density $\rho_0/(1+h)^3$. The internal potential of a homogeneous sphere is 
\begin{equation}
    V(r) = \frac{2}{3}\pi G \rho (r^2 - 3a^2),
\end{equation}
where $r$ is the radius at which the potential is measured. It is most useful to calculate the potential at the referential radius $r$. In this case the physical radius is then $(1+h)r$ and the potential is given by
\begin{equation}
    V(r) = \frac{2}{3} \pi G \frac{\rho_0}{(1+h)} (r^2 - 3a^2).
\end{equation}
This is the potential that will be used for benchmarking. 

\begin{figure}
    \includegraphics[width=\columnwidth]{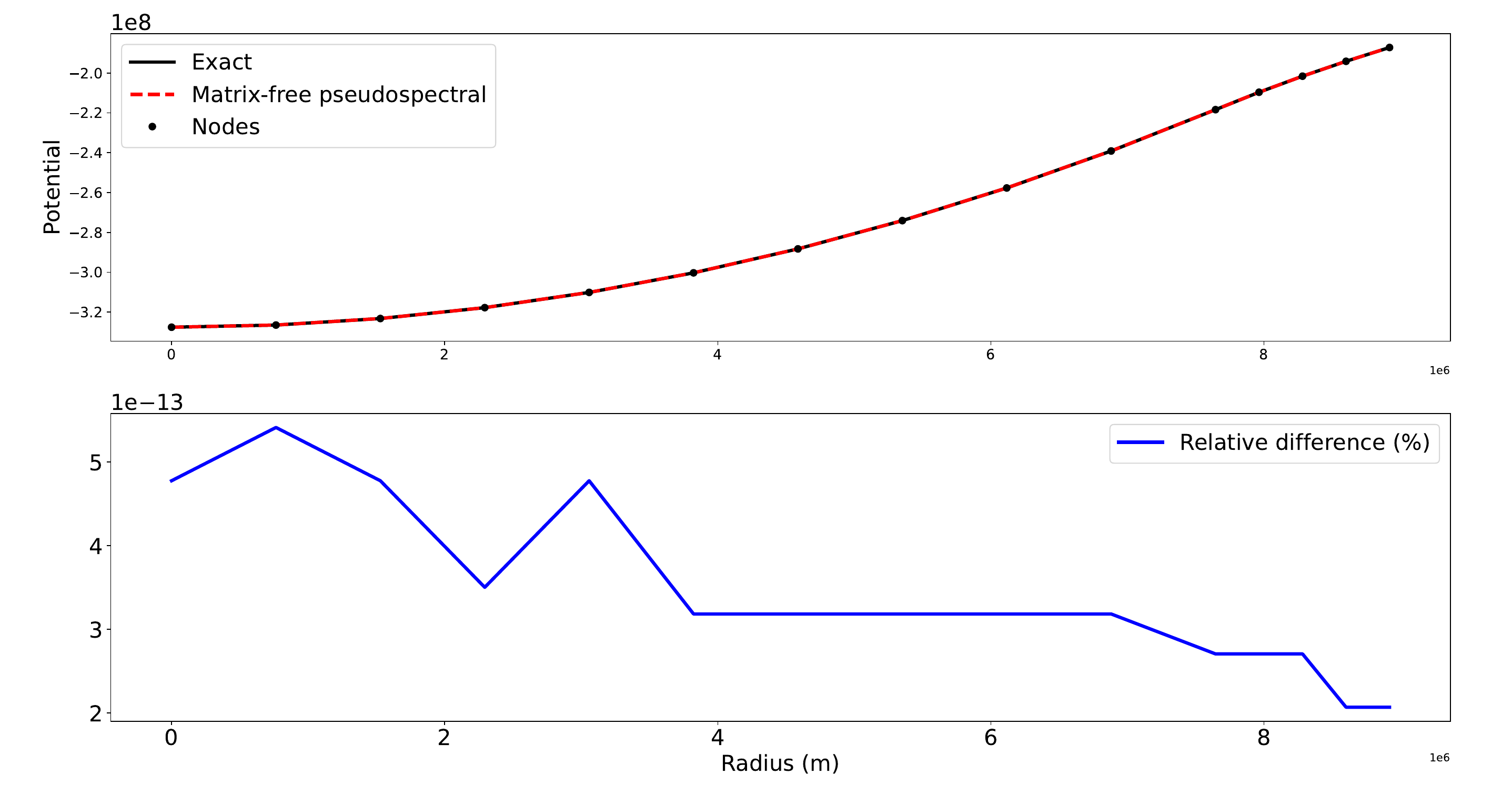}
    \caption{Comparison of potential at the physical radius, for a homogeneous sphere with referential radius equal to that of the Earth ($a$) and referential density the average for the Earth ($\rho$). The radial mapping is $\bs{\xi} = 1.2 \bx$ and the relative error is calculated against the exact solution for a sphere with radius $1.2a$ and density $\rho/1.2^3$.}
    \label{fig:bench1}
\end{figure}

A comparison of the two solutions is given in~\figref{fig:bench1}. The level of accuracy in the conjugate gradient solution was $10^{-12}$ with a spectral element method of $5$th order. As can be seen from the plot, the relative error is of this order and close to machine precision. Furthermore, the number of elements used and the maximum order of the Lagrange polynomials is relatively low, so it is reassuring to have such an accurate computation even in this scenario.

\subsubsection{Physically homogeneous spherical planet with mapping}
The second benchmark shown is for a spherically symmetric homogeneous planet as well. However, we choose a mapping that is not radially symmetric. The referential model is calculated from the physical model during instantiation, using the Jacobian of the mapping and the physical density. 

\begin{figure}
    \includegraphics[width=\columnwidth]{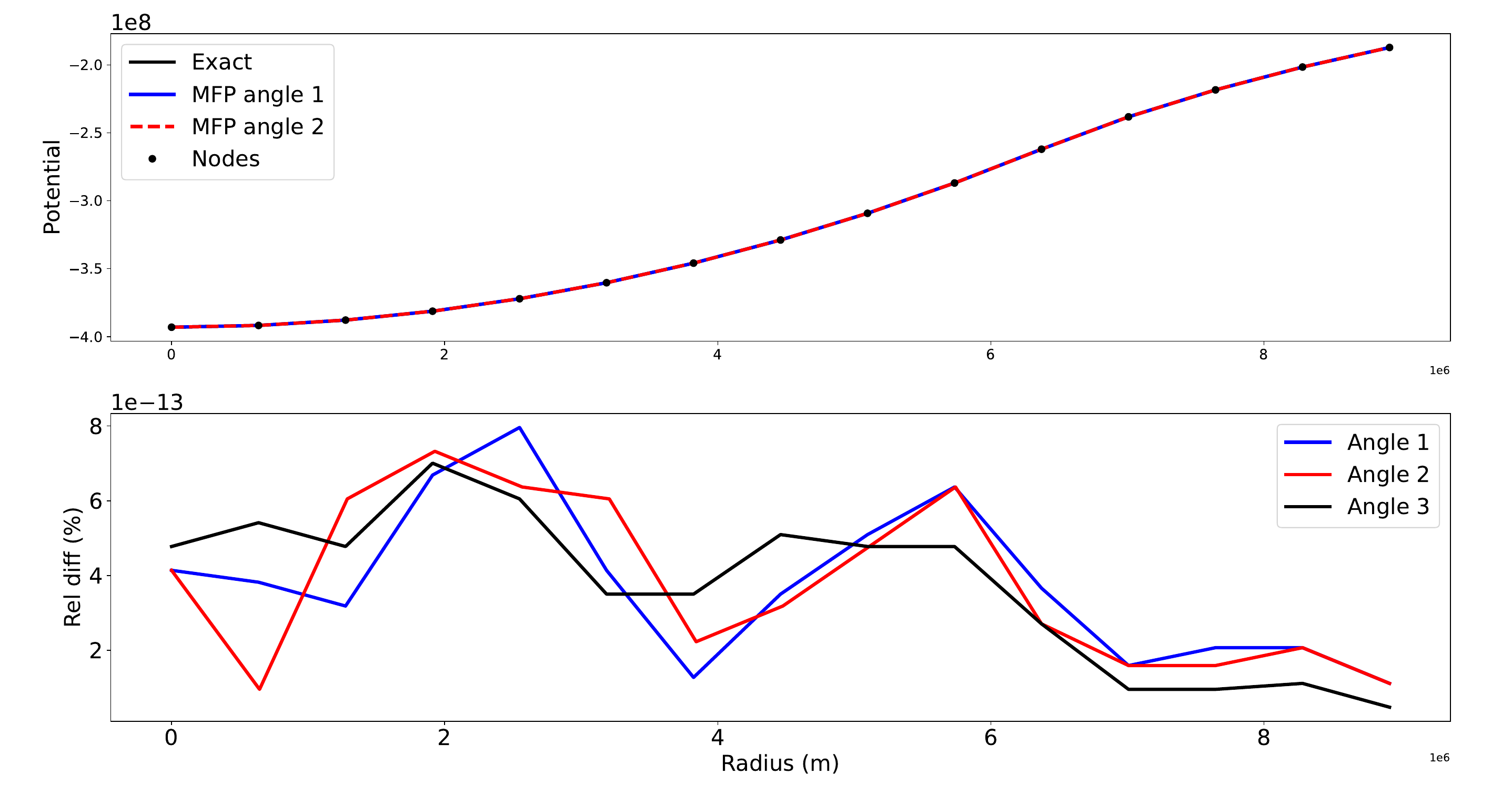}
    \caption{The radial mapping is $\bs{\xi} = 0.2\sin\theta \cos \theta \sin \varphi r(1-r) \bx + \bx$ and the relative error is calculated against the exact solution for a sphere with radius $a$ and density $\rho$. $\lmax=10$ and the order of the interpolation polynomials is eight.}
    \label{fig:bench2}
\end{figure}

The physical planet is of radius $a$, density $\rho_0$, which is homogeneous. The chosen mapping is $\bs{\xi} = 0.2\sin\theta \cos \theta \sin \varphi r(1-r) \bx + \bx$. The result from this calculation at three different combinations of $(\theta,\varphi)$ is shown in~\figref{fig:bench2}.

\subsubsection{Error tolerance}
It is often found that the error is greatest near the origin. The reason for this is that the mapping must decrease to zero as the radius approaches zero. Representation of the matrix $\mbf{a}$ can get difficult in this regime, which can increase the error in the solution. In particular, in the calculation of $\mbf{a}$, one is required to divide by $r$. Consequently, if there is some small mapping close to zero radius, then numerical rounding becomes much more significant. This is an unavoidable result of the method used. However, one can choose a radius $r_l$ below which the mapping is set to zero. This is straightforward to implement and imposes no restrictions on the generality of the method. This has not been done in the benchmarks given above, as can be seen it has not impacted the accuracy of the solution. It may, however, be considered best practice to implement a radius below which the mapping is zero.

\subsection{Forward sensitivity}
\subsubsection{Classical boundary perturbation theory}
As a prerequisite for discussing the benchmarks for the forward sensitivity calculation, we will reprise classical boundary perturbation theory. This discussion is intended to be brief and concerns only those aspects of the theory that are relevant here, for a fuller description please see, e.g.~\citeA{Dahlen_Tromp_1998}. 

The description we use here is Eulerian, i.e. the referential body is the same as the physical body. We use a notation consistent with the notation used in considering perturbations in the referential method. We consider a perturbation to the body whereby a particle, formerly at $\bx$, is displaced by $\delta \bs{\xi}$, i.e. $\mathbf{r}(\bx) = \bx + \delta\bs{\xi}(\bx)$ is the new position of the particle. The density and gravitational potential are consequently perturbed as well, i.e. $\rho(\mathbf{r}) \rightarrow \rho(\mathbf{r}) + \delta\rho(\mathbf{r})$, $\phi\rightarrow  \phi + \delta\phi$. Conservation of mass requires that $\delta\rho = -\nabla \cdot (\rho \bs{\xi})$. The perturbed Poisson equation is $\nabla^2 \delta\phi = 4\pi G \delta\rho$. Obtaining the boundary conditions requires either tedious algebra or the use of a pill box argument but ultimately one finds $\hat{\bn} \cdot [\nabla \delta\phi + 4\pi G \rho \delta\bs{\xi}]_{-}^{+} = 0$. The problem that requires solution is, therefore,
\begin{align}
    \nabla^2 \delta\phi & = -4\pi G \nabla\cdot(\rho \delta \bs{\xi}), \\
    [\delta\phi]_{-}^{+} & = 0, \\
    \hat{\bn} \cdot [\nabla \delta\phi + 4\pi G \rho \delta \bs{\xi}]_{-}^{+} & = 0.
\end{align}
It is trivial to show that the corresponding weak form is given by
\begin{equation}
    \int_{\mb} \nabla \delta\phi \cdot \nabla \bar{\chi} dV + \sum_{lm} (l + 1) b \delta\phi_{lm}(b) \bar{\chi}_{lm} = -4\pi G \int_{\oplus} \rho \delta \bs{\xi} \cdot \nabla \bar{\chi}\dd V. \label{eqn:bptheoryclassical}
\end{equation}
The left hand side of this weak form is recognisable as the weak form of the Poisson equation for $\delta \phi$, with an identity mapping. The identity mapping arises from the Eulerian approach. To facilitate a direct comparison with the result obtained from~\ref{sec:fullpertforce}, one requires a Lagrangian (or referential) description. The reader may question why the Lagrangian, or referential description was not used throughout. The calculations for the classical perturbation theory are more straightforward in an Eulerian framework and the solution is found more easily by simply adding the advective term in, i.e. $\delta \phi\rightarrow \delta \phi + \delta\bs{\xi} \cdot \nabla \phi$. We use this form to benchmark the boundary perturbation results in the case of an initially spherical (physical) planet. The force term is calculated using the canonical components approach and is given by
\begin{equation}
    \text{RHS}= -4\pi G \int_{0}^{b} \left\{r^2 s_{lm}^0(r)h_n'(r)  + \Omega_l^0 r h_{np}(r) [s_{lm}^{-} + s_{lm}^{+}]\right\} \dd r ,
\end{equation}
where $s_{lm}^{\alpha}$ is the component of the $\alpha$-th generalised spherical harmonic of $l$-th degree, $m$-th order and $h_{np}(r)$ is the $p$-th Lagrange polynomial in the $n$-th element.

\subsubsection{Benchmark against classical theory}
\begin{figure}
    \includegraphics[width=\columnwidth]{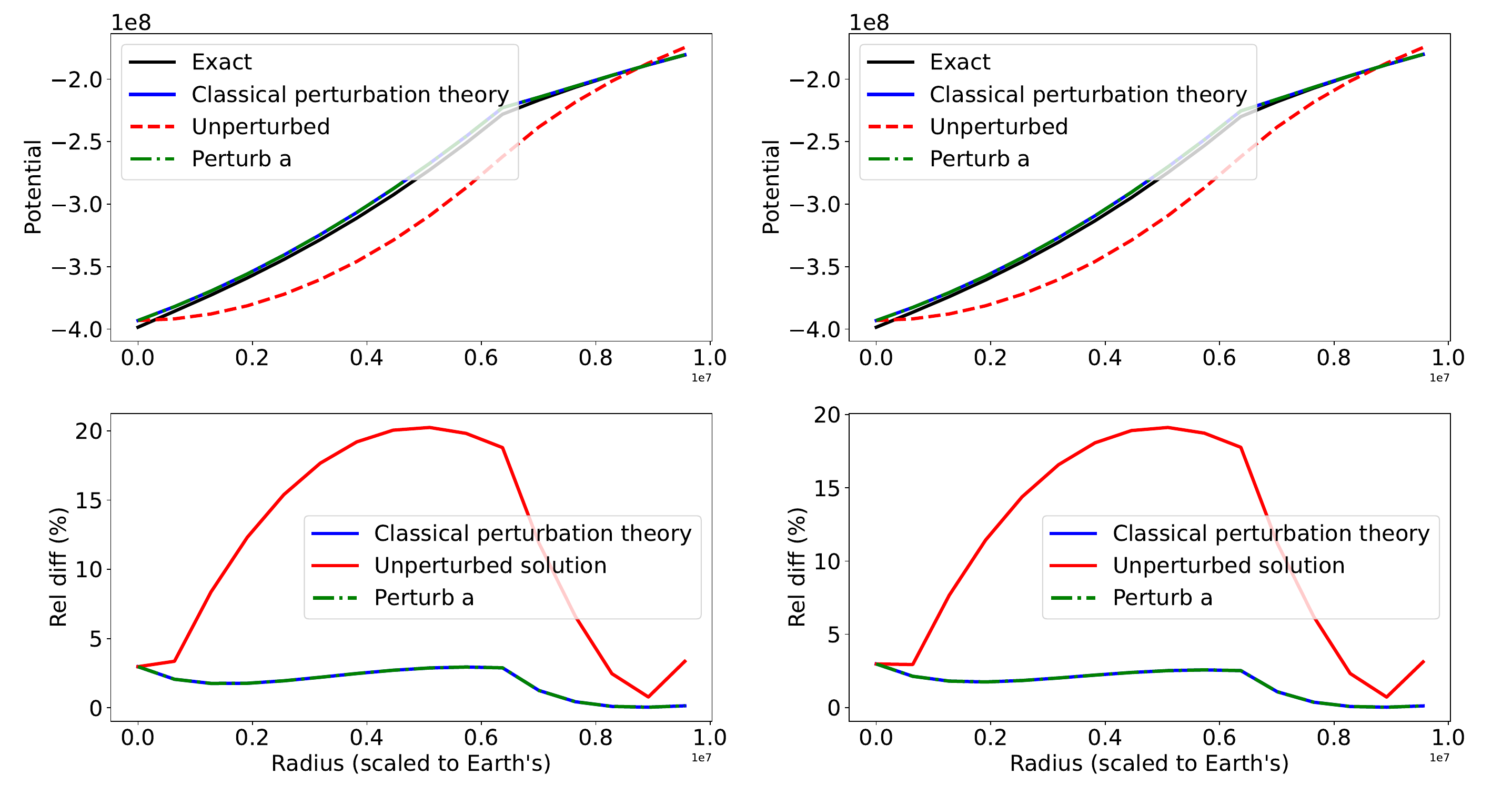}
    \caption{Comparison of the exact, classical perturbation theory and referential perturbation theory. The radius is the scaled referential radius. The unperturbed mapping is the identity and the perturbation is $\delta \bs{\xi} = 0.2 \cos \theta \bx$. The relative error is calculated against the exact solution with the mapping $\bs{\xi} = 1.2 \cos \theta \bx$.}
    \label{fig:benchperturbclassical}
\end{figure}

The unperturbed planet is a homogeneous sphere. The perturbation that we choose for the benchmark is $\delta \bs{\xi} = 0.2 \cos \theta \bx$. The comparison of the ``new'' and classical perturbation methods is given in~\figref{fig:benchperturbclassical}. The chosen perturbation is quite large. However, the results including the perturbation theory match the exact results considerably better than the unperturbed solution. It is noteworthy that, at least in this very simple case, the classical and referential perturbation results are identical.

\section{Sensitivity kernels}
\subsection{Adjoint sensitivity kernels}
\subsubsection{Spherical harmonic component}
In this example we consider the sensitivity kernel of density for an arbitrary spherical harmonic component of the potential at a radius $b$. The planet is chosen to be geometrically spherical, with radius $a < b$. As a matter of intellectual curiosity we demonstrate that the sensitivity kernel we obtain is identical to that which can be derived via more standard methods. If we choose the ball radius to be $b$ then the objective functional in the referential formulation is
\begin{equation}
    Q = b^{-2}\int_{\partial \mb} \overline{Y}_{lm} \zeta \dd S,
\end{equation}
implying that $\tilde{f}(\zeta) = b^{-2}\overline{Y}_{lm} \zeta$. Consequently the weak form for the adjoint equation is
\begin{equation}
    \int_{\mb} \left<\mbf{a} \nabla \chi, \nabla\bar{\eta} \right> \dd^3 \mathbf{x} + \sum_{lm} (l + 1) b \chi_{lm}(b) \bar{\eta}_{lm}(b) = -b^{-2}\int_{\partial \mb}  \overline{Y}_{lm} \bar{\eta} \dd S. \label{eqn:adjointsph}
\end{equation}
As, in this example, we are considering the case of a geometrically spherical planet we know that $\mbf{a} = \mbf{I}$. The second term on the left hand side represents the normal derivative of $\chi$ on the exterior part of the boundary. Thus
\begin{equation}
    \int_{\mb} \left< \nabla \chi, \nabla\bar{\eta} \right> \dd^3 \mathbf{x} - \int_{\partial \mb} \{[\hat{\mathbf{n}} \cdot \nabla \chi]_{\text{E}} - b^{-2}\bar{Y}_{lm} \} \bar{\eta} \dd S =0, \label{eqn:finalweakadjoint}
\end{equation}
where $[\hat{\mathbf{n}} \cdot \nabla \chi]_{\text{E}}$ denotes the normal derivative evaluated exterior to the surface $\partial \mb$. The equivalent strong form of \eqref{eqn:finalweakadjoint} is 
\begin{equation}
    \nabla^2 \chi = 0, \bx \in \mathbb{R}^3, \text{ subject to } [\hat{\mathbf{n}} \cdot \nabla \chi]_-^+ = b^{-2}\bar{Y}_{lm} \text{ on } \partial \mb.
\end{equation}
This equation can be analytically solved by decomposing $\chi = \chi(r) \bar{Y}_{lm}$. This yields the ODE for $\chi(r)$
\begin{equation}
    \chi'' + 2 r^{-1} \chi' - l(l+1)r^{-2} \chi = 0, \text{ subject to } \partial_r \chi|_b + \left(\frac{l+1}{b^3}\right) \chi + 1 = 0.
\end{equation}
The solution of this equation is trivial and we obtain
\begin{equation}
    \chi(r) = -\frac{1}{2l+1} \frac{r^l}{b^{l+1}}.
\end{equation}
Finally this implies that the sensitivity kernel is
\begin{equation}
    \kappa_{\rho} = -\frac{4\pi G}{2l + 1}  \frac{r^l}{b^{l+1}} \bar{Y}_{lm}.
\end{equation}
The radial integral solution of order $lm$, for a geometrically spherical planet, is
\begin{equation}
    \phi_{lm} = -\frac{4\pi G}{2l+1}  \left[r^{-(l + 1)}\int_{0}^r  r'^{2+l} \rho_{lm}(r')\dd r' + r^{l} \int_{r}^{\infty}  r'^{1-l} \rho_{lm}(r')\dd r'\right].
\end{equation}
The radius upon which we measure the spherical harmonic component is greater than the radius of the planet and hence
\begin{equation}
    \phi_{lm}(b) = -\frac{4\pi G}{2l+1} b^{-(l+1)} \int_{0}^r  r'^{2+l} \dd r' \int_{\mathcal{S}^2}  \bar{Y}_{lm} \rho \dd S.
\end{equation}
The sensitivity kernel obtained from the adjoint method is therefore identical to that obtained using the radial integral. 

\subsubsection{Numerical sensitivity kernel}
\begin{figure}
    \includegraphics[width=\columnwidth]{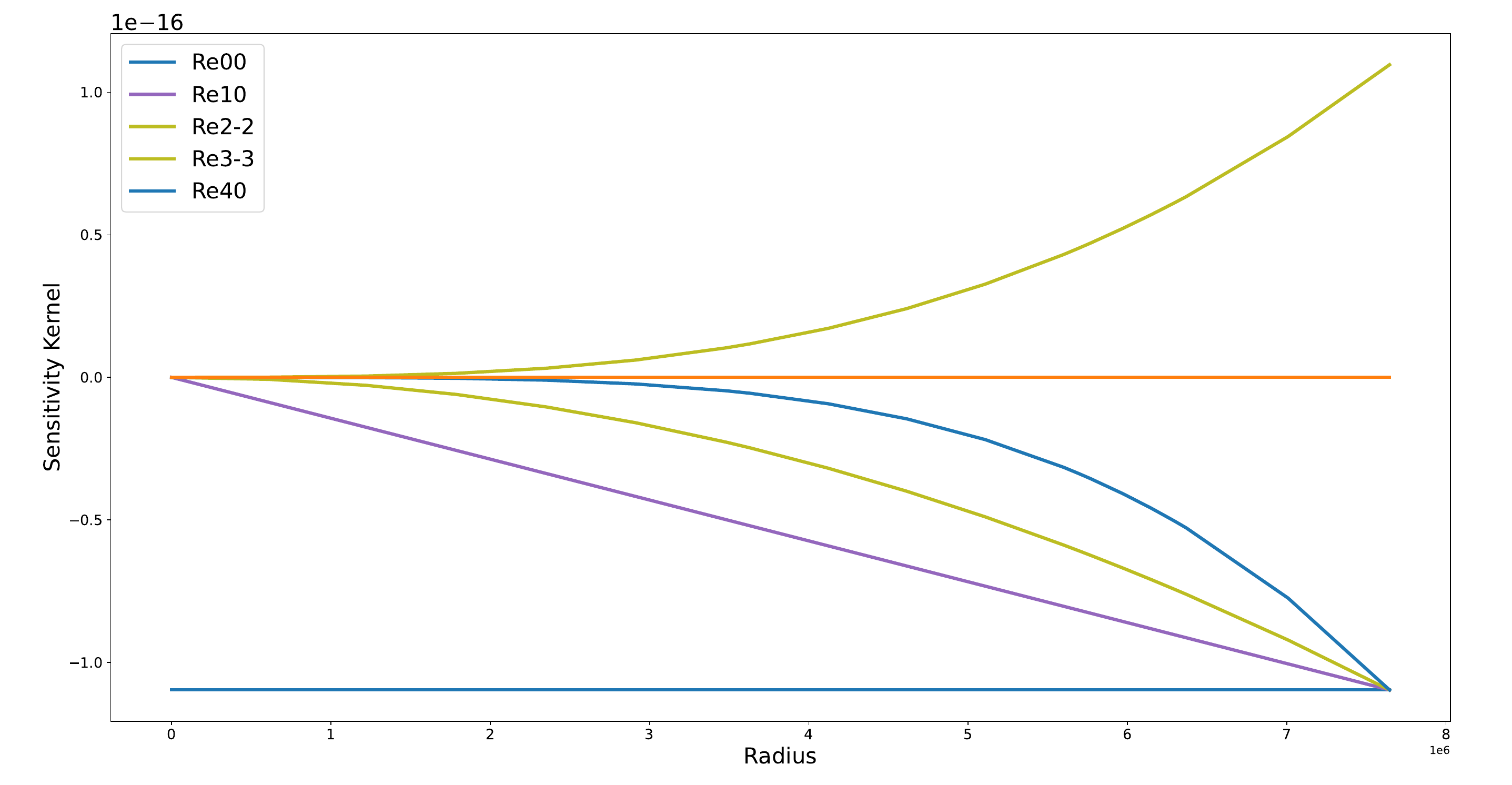}
    \caption{A benchmark of the sensitivity kernel evaluation. The model chosen in which to perform the calculation was a geometrically spherical Earth (with a 1D background of PREM and 3D tomography model S40RTS). The objective functional chosen is a linear combination $f(\zeta) = b^{-2}\zeta [Y_{00} + 3Y_{10} + 5Y_{22} + 7Y_{33} + 9Y_{40}]$. The plot indicates the sensitivity kernel's expression in each spherical harmonic at that radius. Only non-zero sensitivity kernels are shown. }
    \label{fig:sensitivitybench}
\end{figure}

To benchmark the sensitivity kernel we use the exact solution for the spherical harmonic component in a geometrically spherical planet. We plot the sensitivity kernels for multiple spherical harmonic components in~\figref{fig:sensitivitybench}. The chosen objective functional was $f(\zeta) = b^{-2}\zeta [Y_{00} + 3Y_{10} + 5Y_{22} + 7Y_{33} + 9Y_{40}]$. Using the symmetry properties of the spherical harmonics, the sensitivity kernel is
\begin{equation}
    \chi = -\frac{4\pi G}{b}\left[Y_{00}+\frac{r}{b}Y_{10} + \left(\frac{r}{b}\right)^2 Y_{2-2}- \left(\frac{r}{b}\right)^3 Y_{3-3} +  \left(\frac{r}{b}\right)^4 Y_{40}\right].
\end{equation}
The numerically calculated sensitivity kernel is shown in~\figref{fig:sensitivitybench}. It matches exactly the analytically derived sensitivity kernel.

\subsection{Proof of the referential identity for the Poisson problem}
In the Poisson problem the sensitivity kernels for density and deformation gradient were derived. In this appendix we show that they satisfy the referential sensitivity kernel identity. To do this we need to find $\nabla \cdot \mbf{K}_F$. This is given by
\begin{equation}
    \nabla\cdot \mbf{K}_F = \nabla \cdot \left[\left<\mbf{a} \nabla \zeta, \nabla \chi\right>  \mbf{F}^{-T} -    (\mbf{F}^{-T} \nabla\chi)\otimes  (\mbf{a} \nabla \zeta) -   (\mbf{F}^{-T} \nabla \zeta) \otimes (\mbf{a} \nabla \chi)\right].
\end{equation}
Expanding,
\begin{align}
    \nabla\cdot \mbf{K}_F  &= \nabla \cdot\left[\left<\mbf{a} \nabla \zeta, \nabla \chi\right>  \mbf{F}^{-T}\right] -  (\mbf{F}^{-T} \nabla\chi) \nabla\cdot(\mbf{a}\nabla\zeta)-  \nabla (\mbf{F}^{-T} \nabla\chi) \cdot (\mbf{a} \nabla \zeta)\nonumber\\
    & \phantom{=}-  (\mbf{F}^{-T} \nabla \zeta)\nabla\cdot (\mbf{a} \nabla \chi) -\nabla (\mbf{F}^{-T} \nabla \zeta)\cdot(\mbf{a} \nabla \chi).
\end{align}
For reasons that will become clear we will ignore the two divergence terms involving $\mbf{a}\nabla\chi$ and $\mbf{a}\nabla\zeta$ for the moment. We employ index notation which gives
\begin{align}
    T_i &= \partial_l[a_{jk}(\partial_k \zeta)(\partial_j \chi) (F^{-1})_{li}] - (a_{jk}(\partial_k \zeta)) \partial_j [(F^{-T})_{il}\partial_l \chi] - (a_{jk}(\partial_k \chi)) \partial_j [(F^{-T})_{il}\partial_l \zeta]. \label{eqn:zeroterm}
\end{align}
The first term can be expanded using the definition of $\mbf{a}$, and the Piola identity, and is given by 
\begin{align}
    \partial_l[a_{jk}(\partial_k \zeta)(\partial_j \chi) (F^{-1})_{li}] & =  \partial_l[J (F^{-1})_{jm} (F^{-1})_{km} (\partial_k \zeta)(\partial_j \chi) (F^{-1})_{li}], \nonumber\\
    & = J(F^{-1})_{li} (F^{-1})_{jm,l} (F^{-1})_{km} (\partial_k \zeta)(\partial_j \chi) \nonumber\\
    & \phantom{=}+ J(F^{-1})_{li} (F^{-1})_{jm} (F^{-1})_{km,l} (\partial_k \zeta)(\partial_j \chi) \nonumber\\
    & \phantom{=} + a_{jk}(F^{-1})_{li} (\partial_l\partial_k \zeta)(\partial_j \chi) \nonumber\\
    & \phantom{=} + a_{jk}(F^{-1})_{li} (\partial_k \zeta)(\partial_l\partial_j \chi) , \label{eqn:refprooffirstterm}
\end{align}
where a index after a subscripted comma indicates a derivative with respect to that index. We can similarly expand the second and third terms in~\eqref{eqn:zeroterm} to get
\begin{align}
    & (a_{jk}(\partial_k \zeta)) \partial_j [(F^{-T})_{il}\partial_l \chi] + (a_{jk}(\partial_k \chi)) \partial_j [(F^{-T})_{il}\partial_l \zeta] \nonumber\\
    & = a_{jk} (F^{-1})_{li,j} (\partial_k \zeta) (\partial_l \chi) \nonumber\\
    & + a_{jk} (F^{-1})_{li,j} (\partial_k \chi) (\partial_l \zeta) \nonumber\\
    & + a_{jk} (F^{-1})_{li} (\partial_k \zeta) (\partial_j \partial_l \chi) \nonumber\\
    & + a_{jk} (F^{-1})_{li} (\partial_k \chi) (\partial_j \partial_l \zeta). \label{eqn:refproofsecondterm}
\end{align}
It is trivial to see that the third and fourth terms in~\eqref{eqn:refprooffirstterm} and~\eqnref{eqn:refproofsecondterm} cancel, using the symmetry of $\mbf{a}$. Substituting into~\eqnref{eqn:refproofsecondterm} the definition of $\mbf{a}$ we have
\begin{align}
    T_i & = J(F^{-1})_{li} (F^{-1})_{jm,l} (F^{-1})_{km} (\partial_k \zeta)(\partial_j \chi) + J(F^{-1})_{li} (F^{-1})_{jm} (F^{-1})_{km,l} (\partial_k \zeta)(\partial_j \chi) \nonumber\\
    & \phantom{=} - J (F^{-1})_{jm} (F^{-1})_{km}  (F^{-1})_{li,j} (\partial_k \zeta) (\partial_l \chi) - J (F^{-1})_{jm} (F^{-1})_{km}  (F^{-1})_{li,j} (\partial_k \chi) (\partial_l \zeta). \label{eqn:secondlaststepref}
\end{align}
To go further we require an identity for the derivative of $\mbf{F}^{-1}$. In particular, we use $\mbf{F} \mbf{F}^{-1} = I$, take the derivative of the whole expression and ultimately find
\begin{equation}
    (F^{-1})_{lk,m} = -(F^{-1})_{li}(F^{-1})_{jk} F_{ij,m}.
\end{equation}
The first term in~\eqnref{eqn:secondlaststepref} becomes
\begin{align}
    & -J(F^{-1})_{li} (F^{-1})_{jn}(F^{-1})_{pm}F_{np,l} (F^{-1})_{km} (\partial_k \zeta)(\partial_j \chi) \nonumber \\
    & = -J(F^{-1})_{li} (F^{-1})_{jn}(F^{-1})_{pm}(F^{-1})_{km} F_{np,l}  (\partial_k \zeta)(\partial_j \chi).
\end{align}
Similarly the third term in~\eqnref{eqn:secondlaststepref} becomes
\begin{equation}
    J(F^{-1})_{lm} (F^{-1})_{km}(F^{-1})_{jn}(F^{-1})_{pi} F_{np,l}  (\partial_k \zeta)(\partial_j \chi).
\end{equation}
Swapping $p$ and $l$ and using $F_{np,l} = F_{nl,p}$ we find that these two terms cancel. A similar analysis shows that the second and fourth terms in~\eqnref{eqn:secondlaststepref} also cancel. This implies that 
\begin{equation}
    \nabla\cdot \mbf{K}_F^T  = - \nabla\cdot(\mbf{a}\nabla\zeta) (\mbf{F}^{-T} \nabla\chi)- \nabla\cdot (\mbf{a} \nabla \chi) (\mbf{F}^{-T} \nabla \zeta).
\end{equation}
Now, the strong form of the Poisson equation for $\zeta$ is $\nabla\cdot(\mbf{a}\nabla\zeta) = 4\pi G \rho$ and the adjoint equation has the strong form $\nabla\cdot(\mbf{a}\nabla\chi) = 0$ (with appropriate boundary conditions for both). Consequently, we have 
\begin{equation}
    \nabla\cdot \mbf{K}_F^T = -4\pi G \rho \mbf{F}^{-T} \nabla \chi.
\end{equation}
The referential identity is (as the mapping sensitivity kernel is zero)
\begin{equation}
    \rho \mbf{F}^{-T} \nabla K_{\rho} + \nabla\cdot \mbf{K}_F^T = 0.
\end{equation}
As $K_{\rho} = 4\pi G \chi$ we see that this identity does indeed hold.

%
%

\section*{Open Research Section}
The code used to solve the Poisson equation and produce all the figures can be found in https://github.com/adcm2/gplspec. All data used for Phobos' structure was obtained from the supplementary information in~\citeA{willner2014phobos}. Figures were made with Matplotlib version 3.6.3~\cite{caswell2021matplotlib,hunter2007matplotlib}, available under the Matplotlib license at https://matplotlib.org/.

\acknowledgments
AM acknowledges the receipt of a Gates Cambridge scholarship as well as NERC DTP grant.



\bibliography{bibliography}

\end{document}